\documentclass[journal,twocolumn]{IEEEtran}

\makeatletter
\newcommand{\AddInputPath}[1]{%
  \ifx\input@path\@undefined
    \def\input@path{#1}
  \else
    \g@addto@macro{\input@path}{#1}
  \fi
}
\makeatother
\AddInputPath{{../}}

\usepackage{shellesc}

\usepackage{etex}
\usepackage[font=footnotesize,caption=false]{subfig} 

\usepackage{relsize}

\usepackage[dvipsnames,svgnames,usenames]{xcolor}
\usepackage{array}
\usepackage{booktabs,tabularx}
\usepackage{multirow}
\usepackage[per-mode=symbol,detect-mode=true]{siunitx}
\usepackage{graphicx}

\graphicspath{{images/}} 

\usepackage[T1]{fontenc}
\usepackage{textcomp}
\usepackage[utf8]{inputenc}
\usepackage[final]{microtype}
\usepackage{icomma}
\usepackage{xspace}

\usepackage[tbtags]{amsmath}
\usepackage{amssymb,amsfonts,bm}
\usepackage{mathtools} 
\usepackage{dsfont}
\usepackage{mathrsfs}
\usepackage{accents}
\usepackage{empheq}
\usepackage{nccmath}
\usepackage{balance}
\usepackage{setspace}

\usepackage{color}
\usepackage{calc}
\usepackage{tikz}
\usepackage{pgfplots,pgfplotstable}
\usepgfplotslibrary{groupplots}
\usetikzlibrary{matrix} 
\pgfplotsset{compat=1.18} 
\usepackage{pdftexcmds}
\makeatletter
\newcommand{\strequal}[2]{\pdf@strcmp{#1}{#2}==0}
\makeatother

\usepackage[capitalize]{cleveref}
\usepackage{refcount}

\usepackage[inline]{enumitem}
\usepackage{algorithm}
\usepackage{algpseudocode}
\makeatletter
\newcommand{\algmargin}{\the\ALG@thistlm}
\makeatother
\newlength{\whilewidth}
\algdef{SE}[parWHILE]{parWhile}{EndparWhile}[1]
  {\parbox[t]{\dimexpr\linewidth-\algmargin}{%
     \hangindent\whilewidth\strut\algorithmicwhile\ #1\ \algorithmicdo\strut}}{\algorithmicend\ \algorithmicwhile}%

\algnewcommand{\parState}[1]{\State%
  \parbox[t]{\dimexpr\linewidth-\algmargin}{\strut #1\strut}}

\newlength{\elsifwidth}
\algdef{C}[IF]{IF}{parElsIf}[1]
  {\parbox[t]{\dimexpr\linewidth-\algmargin}{%
    \hangindent\elsifwidth\strut\algorithmicelse\ \algorithmicif\ #1\ \algorithmicthen\strut}}%

\makeatletter
\newcommand\fs@spaceruled{\def\@fs@cfont{\bfseries}\let\@fs@capt\floatc@ruled
  \def\@fs@pre{\vspace{.05in}\hrule height.8pt depth0pt \kern2pt}%
  \def\@fs@post{\kern2pt\hrule\relax}%
  \def\@fs@mid{\kern2pt\hrule\kern2pt}%
  \let\@fs@iftopcapt\iftrue}
\makeatother

\usepackage{glossaries}
\usepackage{ifthen}
\usepackage[noadjust]{cite}
\usepackage{multibib}

\usepackage{comment}
\usepackage{todonotes}
\let\legacytodo\todo
\newcommand{\ruggedtodo}[2][]{\tikzexternaldisable\legacytodo[#1]{#2}\tikzexternalenable}
\renewcommand{\todo}[1]{\ruggedtodo[inline]{#1}}

\makeglossaries

\newacronym{sagin}{SAGIN}{space-air-ground integrated network}
\newacronym{fspl}{FSPL}{free-space path loss}
\newacronym{fso}{FSO}{free-space optical communication}
\newacronym{tle}{TLE}{two-line element set}
\newacronym{ai}{AI}{artificial intelligence}
\newacronym{ann}{ANN}{artificial neural network}
\newacronym{jscc}{JSCC}{joint source-channel coding}
\newacronym{raan}{RAAN}{right ascension of the ascending node}
\newacronym{uav}{UAV}{unmanned aerial vehicle}
\newacronym{hap}{HAP}{high-altitude platform}
\newacronym{lap}{LAP}{low-altitude platform}
\newacronym{6g}{6G}{sixth generation}
\newacronym{cgr}{CGR}{contact graph routing}
\newacronym{dtn}{DTN}{delay-tolerant networking}
\newacronym{fl}{FL}{federated learning}
\newacronym{sfl}{SFL}{satellite federated learning}
\newacronym{fo}{FO}{federated optimization}
\newacronym{dl}{DL}{deep learning}
\newacronym{fedavg}{FedAvg}{federated averaging}
\newacronym{dml}{DML}{distributed ML}
\newacronym{ps}{PS}{parameter server}
\newacronym{ml}{ML}{machine learning}
\newacronym{sgd}{SGD}{stochastic gradient descent}
\newacronym{dsgd}{DSGD}{distributed stochastic gradient descent}
\newacronym{isl}{ISL}{inter-satellite link}
\newacronym{gsl}{GSL}{ground-satellite link}
\newacronym{gs}{GS}{ground station}
\newacronym{ecef}{ECEF}{earth-centered, earth-fixed}
\newacronym{eci}{ECI}{Earth-centered inertial}
\newacronym{ofdm}{OFDM}{orthogonal frequency-division multiplexing}
\newacronym{cp}{CP}{cyclic prefix}
\newacronym{los}{LOS}{line-of-sight}
\newacronym{leo}{LEO}{low earth orbit}
\newacronym{meo}{MEO}{medium earth orbit}
\newacronym{gso}{GSO}{geosynchronous orbit}
\newacronym{geo}{GEO}{geostationary}
\newacronym{ntn}{NTN}{non-terrestrial networks}
\newacronym{eo}{EO}{Earth observation}
\newacronym{iot}{IoT}{internet of things}
\newacronym{iort}{IoRT}{internet of remote things}
\newacronym{irs}{IRS}{intelligent reflecting surface}
\newacronym{socp}{SOCP}{second-order cone program}
\newacronym{soc}{SOC}{second-order cone}
\newacronym{dsl}{DSL}{digital subscriber line}
\newacronym{wsee}{WSEE}{weighted sum energy efficiency}
\newacronym{mmwave}{mmWave}{millimeter wave}
\newacronym{dfg}{DFG}{Deutsche Forschungsgemeinschaft}
\newacronym{haec}{HAEC}{Highly Adaptive Energy-Efficient Computing}
\newacronym{hpc}{HPC}{High Performance Computing}
\newacronym{mac}{MAC}{multiple-access channel}
\newacronym{bc}{BC}{broadcast channel}
\newacronym{siso}{SISO}{single-input single-output}
\newacronym{simo}{SIMO}{single-input multiple-output}
\newacronym{miso}{MISO}{multiple-input single-output}
\newacronym{mimo}{MIMO}{multiple-input multiple-output}
\newacronym{af}{AF}{amplify-and-forward}
\newacronym{df}{DF}{decode-and-forward}
\newacronym{cf}{CF}{compress-and-forward}
\newacronym{mwrc}{MWRC}{multi-way relay channel}
\newacronym{dmmwrc}{DM-MWRC}{discrete memoryless multi-way relay channel}
\newacronym{pde}{PDE}{partial data exchange}
\newacronym{fde}{FDE}{full data exchange}
\newacronym{iid}{i.i.d.\@}{independent and identically distributed}
\newacronym{di}{DI} {difference of increasing}
\newacronym{dc}{DC}{difference of convex}
\newacronym{mm}{MM}{mixed monotonic}
\newacronym{mmp}{MMP}{mixed monotonic programming}
\newacronym{awgn}{AWGN}{additive white Gaussian noise}
\newacronym{wgn}{WGN}{white Gaussian noise}
\newacronym{awg}{AWG}{additive white Gaussian}
\newacronym{sic}{SIC}{successive interference cancellation}
\newacronym{snr}{SNR}{signal-to-noise ratio}
\newacronym{sinr}{SINR}{signal to interference plus noise ratio}
\newacronym{inr}{INR}{interference to noise ratio}
\newacronym{zf}{ZF}{zero-forcing}
\newacronym{mrt}{MRT}{maximum ratio transmission}
\newacronym{mmse}{MMSE}{minimum mean square error}
\newacronym{sud}{SUD}{single user decoding}
\newacronym{dof}{DoF}{degrees of freedom}
\newacronym{gdof}{GDoF}{generalized degrees of freedom}
\newacronym{nnc}{NNC}{noisy network coding}
\newacronym{dmn}{DMN}{discrete memoryless network}
\newacronym{csi}{CSI}{channel state information}
\newacronym{pmf}{pmf}{probability mass function}
\newacronym{dmic}{DM-IC}{discrete memoryless interference channel}
\newacronym{ic}{IC}{interference channel}
\newacronym{gic}{GIC}{Gaussian interference channel}
\newacronym{if}{IF}{interference}
\newacronym{ee}{EE}{energy efficiency}
\newacronym{gee}{GEE}{global energy efficiency}
\newacronym{tin}{TIN}{treating interference as noise}
\newacronym{snd}{SND}{simultaneous non-unique decoding}
\newacronym{sd}{SD}{simultaneous decoding}
\newacronym{hk}{HK}{Han-Kobayashi}
\newacronym{rs}{RS}{rate splitting}
\newacronym{rf}{RF}{radio frequency}
\newacronym{pa}{PA}{power amplifier}
\newacronym{lna}{LNA}{low noise amplifier}
\newacronym{lo}{LO}{local oscillator}
\newacronym{adc}{ADC}{analog-to-digital converter}
\newacronym{dac}{DAC}{digital-to-analog converter}
\newacronym{dsp}{DSP}{digital signal processing}
\newacronym{brd}{BRD}{best response dynamics}
\newacronym{br}{BR}{best response}
\newacronym{ne}{NE}{Nash equilibrium}
\newacronym{lhs}{LHS}{left-hand side}
\newacronym{rhs}{RHS}{right-hand side}
\newacronym{ran}{RAN}{radio access network}
\newacronym{qos}{QoS}{Quality of Service}
\newacronym{ngmn}{NGMN}{Next Generation Mobile Networks}
\newacronym{cap}{CAP}{Capacity Adaptation}
\newacronym{bwa}{BW}{Bandwidth Adaptation}
\newacronym{prb}{PRB}{physical resource block}
\newacronym{se}{SE}{spectral efficiency}
\newacronym{tp}{TP}{throughput}
\newacronym{bs}{BS}{base station}
\newacronym{ue}{UE}{user equipment}
\newacronym{mop}{MOP}{multi-objective optimization problem}
\newacronym{gda}{GDA}{generalized Dinkelbach's algorithm}
\newacronym{midcp}{MIDCP}{mixed integer disciplined convex programming}
\newacronym{lp}{LP}{linear program}
\newacronym{brb}{BRB}{branch reduce and bound}
\newacronym{bb}{BB}{branch and bound}
\newacronym{sit}{SIT}{successive incumbent transcending}
\newacronym{oma}{OMA}{orthogonal multiple access}
\newacronym{noma}{NOMA}{non-orthogonal multiple access}
\newacronym{wlog}{w.l.o.g.\@}{without loss of generality}
\newacronym{lsc}{l.s.c.\@}{lower semi-continuous}
\newacronym{usc}{u.s.c.\@}{upper semi-continuous}
\newacronym{kkt}{KKT}{Karush-Kuhn-Tucker}
\newacronym{ptp}{PTP}{point-to-point}
\newacronym{urllc}{URLLC}{ultra-reliable low latency communications} 
\newacronym{cdf}{CDF}{cumulative distribution function}
\newacronym{esa}{ESA}{European Space Agency}

\glsenableentrycount
\makeglossaries

\usetikzlibrary{positioning}
\usetikzlibrary{calc}
\usetikzlibrary{math}
\usetikzlibrary{fit}
\usetikzlibrary{intersections}
\usetikzlibrary{decorations.pathreplacing}
\usetikzlibrary{decorations.markings}
\usetikzlibrary{3d,angles}
\usetikzlibrary{arrows.meta}

\pgfdeclarelayer{background}
\pgfsetlayers{background,main}

\def\mystrut{\vphantom{hg}}

\pgfplotsset{
    legend image with text/.style={
        legend image code/.code={%
            \node[anchor=center] at (0.3cm,0cm) {#1};
        }
    },
}

\usepgfplotslibrary{colorbrewer}

\tikzset{
	small1/.style={fill=DeepPink},
	small2/.style={fill=DeepSkyBlue},
	small3/.style={fill=MediumSpringGreen},
	ps/.style={fill=Gold},
	link/.style = {semithick},
	plane/.style={plane origin={(#1,0,0)}, plane x = {(#1,0,1)}, plane y = {(#1,1,0)}, rotate around y = -9, canvas is plane}
}

\tikzset{
	antenna/.pic={
		\draw[thick] (0,0) -- ++(120:2mm) -- ++(0:2mm) -- cycle -- (0,-1.5mm);
	}
}


\crefname{equation}{}{}
\crefrangeformat{equation}{(#3#1#4)--(#5#2#6)}
\crefmultiformat{equation}{(#2#1#3)}{ and~(#2#1#3)}{, (#2#1#3)}{, (#2#1#3)}
\crefrangemultiformat{equation}{#3(#1)#4--#5(#2)#6}{, #3(#1)#4--#5(#2)#6}{, #3(#1)#4--#5(#2)#6}{, #3(#1)#4--#5(#2)#6}

\undef\mod
\DeclareMathOperator\mod{mod}

\newcommand{\norm}[1]{\ensuremath{\left\lVert #1 \right\rVert}}

\let\vec\bm

\allowdisplaybreaks[3]

\DeclareSIUnit \dBm {dBm}
\DeclareSIUnit \dBW {dBW}
\DeclareSIUnit \dBi {dBi}
\DeclareSIUnit \bpcu {bpcu}

\DeclareFontFamily{U}{mathx}{\hyphenchar\font45}
\DeclareFontShape{U}{mathx}{m}{n}{
      <5> <6> <7> <8> <9> <10>
      <10.95> <12> <14.4> <17.28> <20.74> <24.88>
      mathx10
      }{}
\DeclareSymbolFont{mathx}{U}{mathx}{m}{n}
\DeclareMathSymbol{\bigtimes}{1}{mathx}{"91}

\usepackage{pgfkeys}
    \def\addlegendimage{\csname pgfplots@addlegendimage\endcsname}

\newtheorem{lemma}{Lemma}

\newtheorem{proposition}{Proposition}

\pgfplotscreateplotcyclelist{default}{%
	blue,mark=*\\%
	red,mark=star\\%
	teal,mark=square*\\%
	brown!60!black,mark=otimes*\\%
}

\definecolor{plot1}{RGB}{228,26,28}
\definecolor{plot2}{RGB}{55,126,184}
\definecolor{plot3}{RGB}{77,175,74}
\definecolor{plot4}{RGB}{152,78,163}
\definecolor{plot5}{RGB}{255,127,0}
\definecolor{plot6}{RGB}{166,86,40}
\tikzstyle{fedsatschedule}=[plot1]
\tikzstyle{fedsat}=[plot2]
\tikzstyle{fedisl}=[plot3]
\tikzstyle{fedavg}=[plot4]
\tikzstyle{fedasync1}=[plot5]
\tikzstyle{fedasync2}=[plot6]

\newcolumntype{P}[1]{>{\centering\arraybackslash}p{#1}}

\ifCLASSOPTIONdraftcls
\AtBeginEnvironment{figure}{}
\fi

\AtBeginEnvironment{algorithmic}{\footnotesize}

\title{On-board Federated Learning for Satellite Clusters with Inter-Satellite Links}
\author{
    Nasrin~Razmi,~\IEEEmembership{Graduate Student Member,~IEEE},
	Bho~Matthiesen,~\IEEEmembership{Member,~IEEE},\\
	Armin~Dekorsy,~\IEEEmembership{Senior~Member,~IEEE},
	and~Petar~Popovski,~\IEEEmembership{Fellow,~IEEE}%
			
		\thanks{
			N.~Razmi, B.~Matthiesen, and A.~Dekorsy are with the Gauss-Olbers Center, c/o University of Bremen, and the Department of Communications Engineering, University of Bremen, 28359 Bremen, Germany (e-mails: \{razmi,matthiesen,dekorsy\}@ant.uni-bremen.de).
			P.~Popovski is with the Department of Electronic Systems, Aalborg University, 9100 Aalborg, Denmark (e-mail: petarp@es.aau.dk). P.~Popovski is also holder of the U~Bremen Excellence Chair in the Department of Communications Engineering, University of Bremen, 28359 Bremen, Germany.
		}%
		\thanks{
			This work is supported in part by the German Research Foundation (DFG)
			under Germany's Excellence Strategy (EXC 2077 at University of Bremen, University Allowance).
		}
\thanks{
			 Part of this work was presented at the IEEE International
Conference on Communications (ICC 2022), Seoul, South Korea, May,
2022 \cite{razmi2021board}.
		}		
		
	}

\begin{document}
\bstctlcite{IEEEexample:BSTcontrol}

\maketitle

\begin{abstract}
The emergence of mega-constellations of interconnected satellites has a major impact on the integration of cellular wireless and non-terrestrial networks, while simultaneously offering previously inconceivable data gathering capabilities.
This paper studies the problem of running a federated learning (FL) algorithm within low Earth orbit satellite constellations connected with intra-orbit inter-satellite links (ISL)%
, aiming to efficiently process collected data in situ.
Satellites apply on-board machine learning and transmit local parameters to the parameter server (PS). The main contribution is a novel approach to enhance FL in satellite constellations using intra-orbit ISLs. The key idea is to rely on predictability of satellite visits to create a system design in which ISLs mitigate the impact of intermittent connectivity and transmit aggregated parameters to the PS. We first devise a synchronous FL, which is extended towards an asynchronous FL for the case of sparse satellite visits to the PS. An efficient use of the satellite resources is attained by sparsification-based compression the aggregated parameters within each orbit. Performance is evaluated in terms of accuracy and required data transmission size.
We observe a sevenfold increase in convergence speed over the state-of-the-art using ISLs, and $10\times$ reduction in communication load through the proposed in-network aggregation strategy.
\end{abstract}


\section{Introduction}

Satellite constellations have been an essential component of modern communication and remote sensing systems for decades.
Recent advances in satellite technology and the emergence of interconnected mega constellations in \cgls{leo}, are revolutionizing the way we collect and process data from space \cite{sweeting2018modern,9217520,Abdelsadek2023}. Unlike the previous cellular generations that were exclusively focused on terrestrial networks, mega-constellations and Non-Terrestrial Networks (NTN) are seen as the integral part of 5G and the upcoming 6G wireless systems \cite{Mozaffari6GSky}. 
Consisting of thousands of satellites, these constellations have the potential to process vast amounts of data{\color{black}, e.g., high-resolution hyperspectral images}. Conventional central processing of this collected {\color{black}data} involves significant challenges, including communication delays, limited bandwidth and storage, as well as data ownership concerns \cite{Chen2022,matthiesen2022federated}. The on-board intelligence of satellites increases steadily \cite{Izzo2022,fontanesi2023artificial,Abdelsadek2023}, {\color{black}e.g., PhiSat-1 of \cgls{esa} (ESA) mission,} pushing towards 
in-orbit data preservation and learning to conserve bandwidth and energy, avoid overloading of \cglspl{gsl}, and enable native \cgls{ai} in space.

\cGls{sfl} has emerged as a promising solution to address these challenges \cite{matthiesen2022federated,WCL_fedsat}, as an instance of distributed \cgls{ml} that enables satellites to collaboratively learn a  model without exchanging raw data. With \cgls{fl}, each satellite trains a local model with its own data and sends only the updated model parameters to be aggregated at a central \emph{\cgls{ps}}. FL has the potential to reduce both communication cost and training delay. {\color{black}Nonetheless, the intermittent connectivity between satellites and the \cgls{ps} introduce extended delays when implementing conventional \cgls{fl} in satellite constellations.
The first step towards \cgls{sfl} was made in \cite{WCL_fedsat}, where each satellite acts as an individual collaborator towards the \cgls{ps}, located within a terrestrial \cgls{gs}. 
Each} {\color{black}\cgls{leo}} satellite has only a very short communication window  per orbital period towards the terrestrial \cgls{ps}. This, as well as the fact the link between {\color{black}a} \cgls{gs} and a satellite vanishes behind the horizon for several hours after a few orbital periods, leads to a connectivity bottleneck that severely inhibits convergence speed of a plain \cgls{fl}. Thus, instead of using conventional synchronous \cgls{fl}, \cite{WCL_fedsat} advantageously uses the sporadic, but predictable, satellite connectivity to roll out an asynchronous aggregation.  

Newer satellites, especially within the context of mega constellations \cite{Radhakrishnan2016,Lee2021,Li2022}, rely increasingly on \cglspl{isl} and multi-hop routing. In this paper, we consider a \cgls{sfl} setup with \cglspl{isl} {\color{black} to facilitate the efficient  implementation of both, synchronous and asynchronous \cgls{sfl}. The focus is on scenarios with connectivity} between adjacent satellites within the same {\color{black}circular} orbital plane.
{\color{black} In this case, {\color{black}these} connected satellites have stable relative positions, resulting in an approximately fixed distance from each other and, thus, in stationary link budgets. This is in stark contrast to links across {\color{black}different circular} orbital planes. These inter-orbit \cglspl{isl} are, in the best case, constantly changing in distance and, in the worst case, have a very short lifespan \cite{Bhattacherjee2019,Chaudhry2021,9327501,9217520}. However, we explicitly note that the focus on intra-orbit \cglspl{isl} does not exclude scenarios where additional inter-orbit \cglspl{isl} are available. Indeed, the current work is directly applicable to those scenarios and, in some cases, it might be even preferable to employing all available links, as this will likely result in a considerably higher orchestration complexity.}

{\color{black}A }direct implementation of multi-hop routing leads to network traffic growing quadratically in the number of satellites and a high load on links towards the \cgls{ps}, as each client update will be treated as a common unicast message. Leveraging the structure of \cgls{fl} traffic along with in-network aggregation \cite{Fasolo2007}, communication can be limited to a single outgoing message per satellite and a global iteration during the aggregation phase. Moreover, most of these transmissions use intra-orbit \cglspl{isl} instead of the more challenging \cgls{ps} link, resulting energy saving and reduced communication load at the \cgls{ps}.

\begin{figure}
	\centering
	\includegraphics[clip,trim=0 1.1cm 0 0,width=\linewidth]{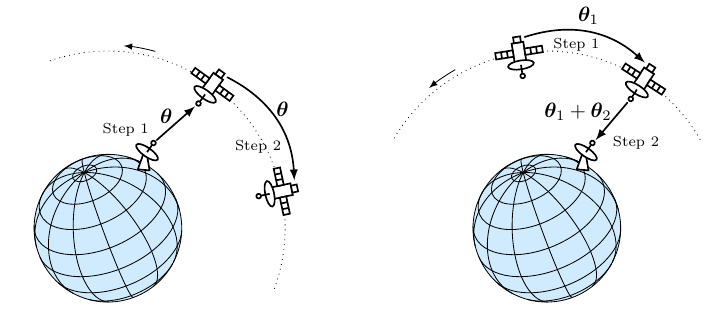}
	\caption{Global model distribution (left) and collection of local updates (right) using intra-orbit inter-satellite communication.}
	\label{fig:Introduction_FedISL}
\end{figure}

\cref{fig:Introduction_FedISL} illustrates the key idea through an example of connectivity bottleneck in \cgls{sfl} using
two satellites at \SI{550}{\km} altitude, spaced \SI{45}{\degree} apart within the same orbital plane.
The \cgls{ps} is located in a \cgls{gs} and both satellites are participating in a \cgls{fl} procedure with their local data sets.
A single pass over the \cgls{gs}, i.e., the time the satellite and the \cgls{gs} can communicate, is less than 10 minutes, while a single orbital period is 95~minutes.
Assume that the computation of this update takes $15$ minutes. Let one of the satellites be selected by the \cgls{ps} to compute an update to the global model. The satellite will collect the current global model on the first pass and return the result on the next pass. This incurs a delay of roughly one orbital period, which is more than one hour in excess of the computation time.
If both satellites are supposed to compute an update then, 
upon the first pass, each satellite will collect the current model version and start computing. The first satellite will pass the \cgls{ps} for the second time and deliver the update. However, in a synchronous \cgls{fl} procedure, the \cgls{ps} will not update the global model until all updates are received. Hence, the first satellite will not have a new model version available to iterate upon during the offline period following the second pass. Differently from this, when \cglspl{isl} are available, the first satellite can transmit this global model to the second satellite directly after receiving it. Then, both satellites compute the update in parallel and the second satellite can collect and deliver both updates to the \cgls{gs} during its first pass. This is illustrated in \cref{fig:Introduction_FedISL}. The time for a single global iteration is reduced from over one orbital period to less than 20 minutes by this approach, while the traffic at the \cgls{ps} is reduced by \SI{50}{\percent}. Implementing this idea requires careful system design and route planning, relying upon the inherent determinism of satellite movement.

The objective of this paper is to present a system design for \cgls{sfl}
within a \cgls{leo} satellite mega-constellation in which the satellites within the same orbital plane are connected via \cglspl{isl} to the adjacent nodes, forming a ring network. Each satellite has the capability to communicate with an out-of-orbit entity, such as \cgls{gs}, that orchestrates the training process. The external orchestration is needed as the satellites are not necessarily able to communicate across orbital planes; at the same time it creates a connectivity bottleneck. The contributions of this paper are:
\begin{itemize}
	\item Design of a distributed system for \cgls{sfl}, supporting client clustering, synchronous and asynchronous orchestration, and consistent decentralized routing decisions.
	\item Development of a communication scheme for \cgls{sfl} that takes advantage of intra-orbit \cglspl{isl}. Due to the co-design of predictive routing and in-network aggregation, the convergence time is reduced markedly, while not increasing the communication load. \color{black}{We have also devised a failure handling procedure.}
	\item Extension of the proposed communication scheme to accommodate gradient sparsification and in-network aggregation for bandwidth-efficiency. This includes the development of a novel estimator for the number of non-zero elements in the sum of sparse vectors.
	\item An effective method to prevent biased solutions for asynchronous aggregation in satellite constellations connected with \cglspl{isl}. This is necessary after improving connectivity with \cglspl{isl}, as simple opportunistic scheduling can result in a small subset of clusters dominating the training process for several hours. 
		%
      \item We have evaluated the performance of the proposed system in several setups. The numerical results highlight a major increase in convergence speed (${\sim}7$ times) due to the use of \cglspl{isl} and ${\sim}10$ times reduction of communication load based on our in-network aggregation approach. 
\end{itemize}

We remark that the proposed system is agnostic to the actual \cgls{fo} procedure, as long as it supports partial aggregation as discussed in \cref{sec:routing:agg}. Hence, the system performance can be further improved using conventional fine-tuning of \cgls{fo} algorithms and \cgls{ml} model-specific hyperparameters \cite{wang2021field}. Finally, we
note that the primary performance metric in this paper is the convergence time measured by a (simulated) wall clock. Conventional metrics for \cgls{fl} algorithms are model test accuracy versus the number of global iterations or the number of gradient computations. This is sensible as the focus in these studies is on improved computational efficiency. Instead, the key challenge that sets \cgls{sfl} apart from conventional scenarios is the connectivity bottleneck implied by the laws of orbital mechanics, potentially leading to extensive delays between iterations. Success in overcoming these obstacles is best measured in the number of global iterations the distributed system can manage within a certain period. As with any communication system, another important metric is bandwidth-efficiency, especially in \cglspl{gsl}.

\subsection{Related Work}

A detailed technical model of \cGls{sfl} was introduced {\color{black}in} \cite{WCL_fedsat}, while \cite{Chen2022} considered it in a more general context of satellite-based computing networks. A broad overview of the different scenarios encountered in \cgls{sfl}, together with a discussion of the technical challenges for each scenario, is presented in \cite{matthiesen2022federated}. The technical details of several of the ideas sketched in \cite{matthiesen2022federated} are provided in this paper. Aerial \cgls{fl} \cite{Pham2022} is a setup that is complementary to \cgls{sfl}, in which the \cgls{ps} is operated within {\color{black}satellites or aerial stations} to orchestrate a planet-side \cgls{fl} process{\color{black}. It also contains scenarios where the aerial stations act as clients participating in the training process. The focus of \cite{Pham2022} is, however, on using \cgls{ntn} as access network for terrestrial \cgls{fl} nodes}.
Closely related to that are the satellite-assisted \cgls{iort} system architectures considered in \cite{9982621} and \cite{Uddin2022}{\color{black}, where satellites serve as \cgls{ps} and access network, respectively.}

A {\color{black}coarse} classification of current work on \cgls{sfl} can be made based on the presence and usage of \cglspl{isl}.
The model in \cite{WCL_fedsat} considers the case without \cglspl{isl}, identifies the connectivity bottleneck, and proposes a satellite-specific asynchronous \cgls{fl} algorithm as solution.
This work is extended in \cite{razmi2022scheduling} with a scheduling algorithm that exploits the inherent determinism of satellite trajectories.
{\color{black} This scheduler can be combined with the clustered approach to \cgls{sfl} presented here. Indeed, while the current work aims at improving worker-\cgls{ps} connectivity to facilitate faster convergence and, optimally, synchronous orchestration, the algorithm in \cite{razmi2022scheduling} focuses on reducing staleness by leveraging on the predictable connectivity.}
Another scheduling approach based on buffered asynchronous \cgls{fl} is proposed in \cite{so2022fedspace}, aiming to balance local model staleness and idle times. Lacking a thorough benchmark against the state-of-the-art, the gain of the complicated scheduling algorithm in \cite{so2022fedspace} remains an open question. Staleness in asynchronous \cgls{sfl} is further investigated in \cite{wu2023fedgsm}, which introduces an asynchronous update rule based on the notion that staleness effects in \cgls{sfl} are similar in consecutive training epochs. In \cite{10092567}{\color{black},} the connection bottleneck is tackled by combining synchronous orchestration with a dynamic aggregation rule that ignores stragglers. However, the primary reason for the feasibility of synchronous terrestrial orchestration is the usage of multiple geographically distributed \cglspl{gs} that act as a distributed \cgls{ps}. While the authors observe correctly that latency between \cglspl{gs} is small compared to \cglspl{gsl}, mechanisms to either ensure consistency between \cglspl{gs} or a hierarchical \cgls{fl} approach would be required in a practical system implementation of \cite{10092567}. 

An alternative means to improving connectivity is the usage of \cglspl{isl} instead of a distributed \cgls{ps}, as proposed in the conference version of this paper \cite{razmi2021board}. The communication strategy from \cite{razmi2021board} is adopted in \cite{elmahallawy2022fedhap} and combined with a distributed \cgls{ps} implemented within interconnected {\color{black}\cglspl{hap}}. This is extended in \cite{Elmahallawy2022a} to asynchronous aggregation with multiple {\color{black}\cglspl{hap}}. A modified version of \cite{razmi2021board} is proposed in \cite{elmahallawy2023optimizing}, consisting of a decentralized implementation of predictive routing, which might lead to inconsistent routing decisions, and the absence of incremental aggregation (see \cref{sec:routing:agg}), which leads to a quadratic traffic growth within each orbital plane. 
%
\cgls{fl} in a {\color{black} fully connected, ultra-dense satellite constellation, employing both intra- and inter-orbit \cglspl{isl},} is considered in \cite{Chen2023}.
{\color{black} There, only satellites within close vicinity of the \cgls{gs} are participating in each epoch of the training process.}
A decentralized learning system, without \cgls{ps}, leveraging inter- and intra-orbit \cglspl{isl} is proposed in \cite{Wu2022}. Decentralized learning in \cgls{leo} satellite constellations under very realistic satellite system assumptions is considered in \cite{ostman2023decentralised} for a semi-supervised classification task.
Finally, \cite{Wang2022} treats \cgls{fl}-aware routing and resource allocation for \cgls{sfl}.

{\color{black}
\subsection{Organization}

The rest of this paper is organized as follows. In \cref{sec: System Model}, we present the system model, which includes models for the constellation, communications, and computation. \Cref{sec:ps} describes the different orchestration approaches at the \cgls{ps} and rigorously defines its operation. In \cref{sec:routing}, the client process is defined and an efficient communication scheme for \cgls{sfl} is developed. \Cref{sec:sparsification} discusses incorporating gradient compression in the communication scheme for increased bandwidth efficiency, using gradient sparsification as an example. Finally, we evaluate the performance of our framework in \cref{sec: Performance Evaluation} and conclude the paper in \cref{sec: Discussion and Conclusions}.
}

\subsection{Notation} Scalars are represented in a normal font $x$, while vectors in bold $\vec{x}$. The Euclidean norm of a vector $\vec{x}$ is $||\vec{x}||$. The angle between two vectors $\vec{x_1}$ and $\vec{x_2}$ is $\angle (\vec{x_1},\vec{x_2})$. 
Sets are denoted by $\mathcal X$, and the cardinality of $\mathcal X$ is $|\mathcal X|$. Removing an element $x_i$ from the set $\mathcal X$ is denoted by $\mathcal X \setminus \{ x_i \}$.
 In a graph $\mathcal G = (\mathcal V, \mathcal E)$ with vertices $\mathcal V$ and edges $\mathcal E$, the neighborhood of any vertex $v \in\mathcal V$ is denoted as $\mathcal N(v)$. If $\mathcal G$ is directed, $\mathcal N^-(v)$ and $\mathcal N^+(v)$ denote the incoming and outgoing neighborhood of $v$, respectively.
 The operators $\Pr(\cdot)$, $\mathds E$, and $I(\cdot)$ are the probability, expected value, and indicator function, respectively. Integer rounding is denoted by $\lfloor{\cdot\rfloor}$ and $\lceil{\cdot\rceil}$.

\section{System Model} \label{sec: System Model}
The constellation has $P$ orbital planes, {\color{black} where orbit $p$, $p\in\{1, \dots, P\}$, contains $K_p$ satellites $\mathcal K_p$ such that $\mathcal K_{p} \cap \mathcal K_{q} = \emptyset$, for all $q \neq p$}. The set of all satellites within the constellation is denoted as $\mathcal K = \bigcup_{p = 1}^P \mathcal K_p = \{ k_{1,1},\dots,k_{P,K_P} \}$, with the total number of satellites $K = \sum_{p=1}^P K_p$.
Each satellite $k$ follows a trajectory $\vec s_k(t)$ around Earth with orbital period $T_p = 2 \pi \sqrt{\frac{a_k^3}{\mu}}$, i.e., $\vec s_k(t) \approx \vec s_k(t+ n T_p)$ for all integer $n$, where $a_k$ is the semi-major axis of satellite $k$ and $\mu = 3.98 \times 10^{14}\,\si{\meter^3/\second^2}$ is the geocentric gravitational constant. For circular orbits, the semi-major axis is $a_k = r_E + h_k$ with $h_k$ being the satellite's altitude above the Earth's surface and $r_E = 6371\,\si{\km}$ the Earth radius. The satellites within an orbital plane $p$ follow the same trajectory and are assigned unique IDs $\mathcal K_p =\{ k_{p,1}, \dots, k_{p,K_p}\}$ such that satellite $k_{p,i+1}$ is behind $k_{p,i}$. If the satellites are distributed equidistantly within the orbital plane, we have
$\vec s_{k_{p,1}}(t) \approx \vec s_{k_{p,2}}(t - T_p / K_p) \approx \vec s_{k_{p,3}}(t - 2 T_p / K_p) \approx \dots$.\footnote{Under the assumption of perfect Keplerian orbits, we can replace '$\approx$' with '$=$' in all statements on trajectories. With real-world orbits being subject to orbital perturbations and station keeping maneuvers, these relations do not hold exactly and we only state them here to introduce notation and some general concepts on an abstraction level suitable for this paper.} Coordinates are in an Earth-centric reference frame.

\subsection{Communication Model}
{\color{black}The number of communication devices per satellite depends on the specific mission requirements. In this paper, we assume e}ach satellite has three communication devices, two of which are for intra-orbit communications. The third one is for communication outside of its orbital plane, either a \cgls{gsl} {\color{black} if the \cgls{ps} is located in a \cgls{gs}} or an \cgls{isl} towards a satellite in another orbit {\color{black} if the \cgls{ps} is located in a satellite.} {\color{black}It is worth noting that if the proposed schemes are applied for satellites equipped with more than three communication devices, only the three required ones are used.} Communication with an Earth-based \cgls{gs} is feasible if the satellite is visible from the \cgls{gs} at an elevation angle $\frac{\pi}{2} - \angle (\vec s_{GS}, \vec s_k(t) - \vec s_{GS}) \ge \alpha_e$, where $\alpha_e$ is the minimum elevation angle \cite{3gpp38.811,9217520} and $\vec s_{GS}$ is the position of the \cgls{gs}. While this condition is satisfied, we assume communication is possible at a fixed rate.\footnote{\color{black}The fixed rate assumption in the out-of-constellation link is made for the sake of simplicity. It has no direct impact on the developed \cgls{sfl} framework and can be relaxed easily to a variable rate if the system supports adaptive coding and modulation.} 
For \cglspl{isl}, communication is feasible if the line of sight is not obstructed by the Earth. With lower atmospheric layers degrading the link quality, a sensible assumption is to consider an {\color{black}\cgls{isl}} to be feasible if it does not {\color{black}enter} the atmosphere below the {\color{black}thermosphere} \cite{Bhattacherjee2019}, starting at approximately \SI{80}{\km} above sea level. This translates to a maximum slant range $d_\mathrm{Th}(t; k_1, k_2) = \sqrt{\norm{\vec s_{k_1}(t)}^2 - r_T^2} + \sqrt{\norm{\vec s_{k_2}(t)}^2 - r_T^2}$ for any two satellites $k_1, k_2$, where $r_T = r_E + \SI{80}{\km}$. For circular orbits, this threshold is the constant
$d_\mathrm{Th}(k_1, k_2) = \sqrt{{(h_{k_1}+r_E)^2}-{r_T^2}}+\sqrt{{(h_{k_2}+r_E)^2}-{r_T^2}}$.
We assume communication at a fixed rate is possible between satellites $k_1$ and $k_2$ if their distance at time $t$ is $d(t; k_1, k_2) \leq d_\mathrm{Th}(t; k_1, k_2)$.

The most stable link usage is to connect each satellite to its two closest orbital neighbors, effectively forming a ring network \cite{9327501}. The two neighbors of satellite $k_{p,i}$ are $\mathcal N(k_{p_i}) = \{ k_{p,i-1}, k_{p,i+1} \}$. Here, the satellite indices $i-1$ and $i+1$ are modulo $K_p$, which is a convention we will adopt throughout this paper until further notice.
Following the previous discussion,
the data rate between any two satellites
$k_1, k_2 \in\mathcal K$ within the constellation is fixed to
$r(t; k_1, k_2) = \rho_{k_1, k_2}$ if $k_2 \in \mathcal N(k_1)$ and communication is feasible, and zero otherwise. For the out-of-orbit communication link, we assume there is a single communication partner of interest, denoted as the \cgls{ps}. Then, the rate function of satellite $k$'s, $k\in\mathcal K$, link towards the \cgls{ps} is similarly defined as $r_\mathrm{PS}(t; k) = \rho_{k,\mathrm{PS}}$ if communication is feasible and zero otherwise.

\subsection{Computation Model}
The satellites within the constellation collaboratively train a \cgls{ml} model from data $\mathcal D$ collected at the satellites. The \cgls{ml} model is known at all satellites and fully defined by its model parameter vector $\vec w \in\mathds R^{n_d}$. The goal is to find a solution to the optimization problem
\begin{equation} \label{eq:opt}
	\min\nolimits_{\vec w}\enskip F(\vec w),
\end{equation}
where the global loss function $F(\vec w) = \frac{1}{D} \sum_{\vec x\in\mathcal D} f(\vec x; \vec w)$, with $D = |\mathcal D|$, measures the performance of the \cgls{ml} model with respect to the data set $\mathcal D$ for a certain set of model parameters $\vec w$ and a potentially nonconvex per-sample loss function $f(\vec x; \vec w)$.

Regarding \cref{eq:opt}, we make two fundamental assumptions: \emph{(i)} the computational resources at the satellites are limited such that a distributed solution of \cref{eq:opt} is necessary; \emph{(ii)}  the data set $\mathcal D$ is distributed across the constellation and communicating this data is not feasible; i.e., each satellite has a local data set $\mathcal D_k$ such that $\mathcal D = \bigcup_{k\in\mathcal K} \mathcal D_k$, which is not shared with any other participants in the training process.
Due to the local data sets assumption and the limited connectivity, this distributed \cgls{ml} scenario is an instance of a \cgls{fl} \cite{mcmahan2017communication}. However, in contrast to a conventional \cgls{fl} setup, here client participation is under central control, the connectivity is mostly deterministic and predictable, and the number of devices is orders of magnitude lower. 

In \cgls{fl}, \cref{eq:opt} is solved iteratively using a modified \cgls{dsgd} procedure.\footnote{While this paper is focusing on \cgls{dsgd}-based optimization, the extension to many other iterative distributed optimization algorithms for training \cgls{ml} models is straightforward.} This algorithm is motivated by the linearity of the gradient and the observation that the objective function is separable as $F(\vec w) = \frac{1}{D} \sum_{k\in\mathcal K} D_k F_k(\vec w)$ with $F_k(\vec w) = \frac{1}{D_k} \sum_{\vec x\in\mathcal D_k} f(\vec x; \vec w)$
and $D_k = |\mathcal D_k|$.
The optimization process is orchestrated by a central \cgls{ps}, which maintains the current iteration of the global model parameters $\vec w$, distributes these to the clients for further refinement, and collects the results. In contrast to conventional \cgls{fl}, we assume that every client participates in every iteration of the solution process due to the relatively small number of clients.
Since satellites within the constellation are only connected to their orbital neighbors, communication across orbital planes is only feasible via the out-of-constellation link. Hence, it is only natural to assume the \cgls{ps} to be either located in a \cgls{gs} or in a satellite outside the constellation. Details of the \cgls{ps} operation will be discussed in \cref{sec:ps}.

\subsubsection{Client Operation}
Each satellite $k\in\mathcal K$ runs a process to handle all application-layer communications related to the \cgls{fl} training.
This procedure will be designed in \cref{sec:routing}. Upon receiving an updated global parameter vector $\vec w^n$, it launches the learning procedure outlined in \cref{alg:Satellite SGD Procedure} in a separate thread for concurrent execution.
This learning procedure then computes an update to $\vec w^n$ based on the local data set $\mathcal D_k$. After initialization in line~\ref{alg:ssgd:init}, the loss function $F_k(\vec w)$ is minimized in $I$ local epochs using permutation-based mini-batch \cgls{sgd} in lines~\ref{alg:ssgd:batchstart}--\ref{alg:ssgd:batchend}.
{\color{black}
	More specifically,
	in each local epoch, satellite $k$ shuffles its data set $\mathcal{D}_k$ randomly and then divides it into mini-batches of size $\mathcal{|B|}$. Subsequently, it performs a gradient step based on the empirical average per-sample loss for each mini-batch, i.e.,
	\begin{equation} \label{SGD}
	\vec{w}_k^{n,i+1} \gets \vec{w}_k^{n,i} - \frac{\eta}{|\mathcal B|} \nabla_{\vec{w}} \left(\sum_{\vec x\in\mathcal B} f(\vec x; \vec w) \right),
	\end{equation}
	where $\eta$ is the learning rate. Instead of directly transmitting the updated model parameters $\vec{w}^{n}_k$, as represented in line 11,
}
the effective gradient $\vec g_k(\vec{w}^{n}_{k})$ is computed in line~\ref{alg:ssgd:eg}. While both {\color{black}representations} $\vec{w}_k^{\color{black}{n}}$ and $\vec g_k(\vec{w}^{n}_{k})$ are theoretically equivalent, the latter is often easier to compress. This is optionally done in line~\ref{alg:ssgd:compress} by calling the procedure \textsc{CompressGradient}, which will be defined in \cref{sec:sparsification}. Without compression, \textsc{CompressGradient} is simply the identity function, i.e., $\bar{\vec g}_{k}(\vec{w}^{n}_{k}) = \vec{g}_k(\vec{w}^{n}_{k})$. In a slight modification of the usual approach, \cref{alg:Satellite SGD Procedure} returns the compressed effective gradient scaled by $D_k$ in line~\ref{alg:ssgd:ret}. 

\begin{algorithm}
	\caption{Satellite Learning Procedure} \label{alg:Satellite SGD Procedure}
	\begin{algorithmic}[1]
		\Procedure{ClientOpt}{$\vec w$}
			\State \textbf{initialize} $\vec{w}_k^{n,0} = \vec{w}^n, \quad i = 0$ \label{alg:ssgd:init}
			\For {$I$ epochs} \label{alg:ssgd:batchstart}
				\Comment $I$ epochs of mini-batch SGD
				\State $\tilde{\mathcal D}_k \gets $ Randomly shuffle $\mathcal D_k$
				\State $\mathscr B \gets $ Partition $\tilde{\mathcal D}_k$ into mini-batches of size $B$
				\For {each batch $\mathcal B\in\mathscr B$}
				\State $\vec{w}_k^{n,i+1} \gets \vec{w}_k^{n,i} - \frac{\eta}{|\mathcal B|} \nabla_{\vec{w}} \left(\sum_{\vec x\in\mathcal B} f(\vec x; \vec w) \right)$
				\State $i \gets i + 1$
				\EndFor
			\EndFor \label{alg:ssgd:batchend}
   \State {\color{black} $\vec{w}_{k}^{n} \gets \vec{w}_{k}^{n,i}$}
			\State $\vec{g}_k(\vec{w}^{n}_{k}) \gets {\color{black}\vec{w}^{n}_{k}}-\vec{w}^{n,0}_{k}$ \label{alg:ssgd:eg}
				\Comment Compute effective gradient 
			\State $\bar{\vec g}_{k}(\vec{w}^{n}_{k}) \gets \Call{CompressGradient}{\vec{g}_k(\vec{w}^{n}_{k})}$ \label{alg:ssgd:compress}
				\Comment {Apply gradient\\ \hspace{4.6 cm}compression (e.g., sparsification)}
			\State \Return $D_k \bar{\vec g}_{k}(\vec{w}^{n}_{k})$ \label{alg:ssgd:ret}
		\EndProcedure
	\end{algorithmic}
\end{algorithm}

Note that the subsequent results do not rely on the explicit implementation of the \cgls{sgd} procedure in lines~\ref{alg:ssgd:batchstart}--\ref{alg:ssgd:batchend}. The only requirement is that the updated model parameters can be incorporated into the global model based on effective gradients using the update rules presented in \cref{sec:ps}. However, we will use the implementation in \cref{alg:Satellite SGD Procedure} throughout this paper.

For the routing procedure developed in \cref{sec:routing}, we will require an accurate estimate of the time to run \cref{alg:Satellite SGD Procedure}. Apart from scheduling delays due to multi-task computing, the runtime directly depends on the number of processor cycles for each operation in \cref{alg:Satellite SGD Procedure}. These are hardware-dependent, deterministic, and can be determined offline before deployment \cite{Miettinen2010}. First, consider a single epoch. The local data set is shuffled and divided into
$\left\lceil \frac{D_k}{B} \right\rceil$ mini-batches. This process takes $c_\mathrm{epoch}$ CPU cycles per sample. Computation of the stochastic gradients requires, in total, $n_d D_k c_\mathrm{s}$ CPU cycles, where $c_\mathrm{s}$ is the number of cycles to process one sample for a single dimension of $\vec w$. Executing the gradient step takes, per mini-batch, $n_d c_\mathrm{step}$ clock cycles.
Thus, one epoch requires a total of
$D_k c_\mathrm{epoch} + n_d D_k c_\mathrm{s} + \left\lceil \frac{D_k}{B} \right\rceil n_d c_\mathrm{step}$
CPU cycles.
After $I$ epochs, a final gradient step taking $n_d c_\mathrm{step}$ cycles is performed to compute the effective gradient. The gradient compression takes an additional $c_\mathrm{compress}$ cycles (see \cref{sec:sparsification}) and is assumed zero if no compression is used.
Passing the result to the communication stack takes, together with other overhead occurring due to, e.g., process setup and termination, a total of $c_\mathrm{os}$ cycles. The runtime of \cref{alg:Satellite SGD Procedure} is

{\small
\begin{equation}
     \label{eq: Time of learning}
	t_l(k)\hspace{-0.3 em}=\hspace{-0.3 em} \frac{I D_k (c_\mathrm{epoch}\hspace{-0.2 em}+\hspace{-0.2 em} n_d c_\mathrm{s})\hspace{-0.2 em}+\hspace{-0.2 em} c_\mathrm{step} n_d \left(\hspace{-0.1 em}I\hspace{-0.1 em} \left  \lceil \frac{D_k}{B} \right\rceil \hspace{-0.3 em}+\hspace{-0.3 em} 1\right)\hspace{-0.3 em}+\hspace{-0.25 em} c_\mathrm{compress}\hspace{-0.25 em}+\hspace{-0.25 em}c_\mathrm{os}}{\nu_k},
\end{equation}}
where $\nu_k$ is the CPU frequency at satellite $k$.

\section{Orchestration of Satellite Federated Learning} \label{sec:ps}
\glsreset{ps}
{\color{black}\cGls{fl}} uses a conventional client-server architecture to orchestrate the training process. While the clients compute the stochastic gradient steps for \cref{eq:opt}, the \cgls{ps} is responsible for aggregating these gradient steps, updating the global model parameters, and distributing the updated parameter vector to the clients. 
In a conventional \cgls{fl} setup, the \cgls{ps} is also responsible for client scheduling, modeled as an uniform sampling of a subset of clients in each global iteration. Given that \cgls{fl} operates on a massive number of clients, this is equivalent to a two-stage \cgls{sgd} step, the first step being a client selection and the second computation. Instead, \cgls{sfl} operates with significantly fewer (orders of magnitude) clients and each client has a much larger share of the total data, exhibiting a non-negligible contribution to the unbiased  model convergence. Hence, it is reasonable that all clients participate in every global iteration.

Connectivity towards the workers is necessary for synchronization of the training process. In \cgls{sfl}, the communication window from a single \cgls{leo} satellite towards a ground-based \cgls{ps} is usually in the order of a few minutes, followed by an offline period due to Earth blockage, ranging from one orbital period up to several hours. As shown in \cite{WCL_fedsat}, the conventional FedAvg operation of collecting all local updates before creating a new global model iteration leads to severe delays in the training process. This bottleneck can be partially mitigated by modifying FedAvg for \emph{asynchronous} operation \cite{WCL_fedsat}.
Compared to synchronous \cgls{fl}, asynchronous \cgls{fl} has a 
slower convergence speed in terms of gradient steps. For ground-orchestrated \cgls{sfl} without \cglspl{isl}, this decrease is greatly outweighed by the reduction in the delay between gradient steps, resulting in much faster overall convergence speed, measured in wall time. Leveraging \cglspl{isl}, the optimal \gls{ps} operation very much depends on the \cgls{ps} location and resulting connectivity patterns \cite{matthiesen2022federated}. We will introduce both orchestration approaches in a unified manner in \cref{sec:routing,sec:sparsification}.
The actual aggregation rule at the \cgls{ps} is easily exchangeable as long as additivity of individual client updates holds. This broadens the contribution of \cref{sec:routing}, as it allows to improve the \cgls{ps} operation while preserving the dense connectivity patterns enabled by \cglspl{isl}.

\subsection{Synchronous Orchestration}
A synchronous \cgls{fl} \cgls{ps}, exemplified by FedAvg \cite{mcmahan2017communication}, repeats the
following steps until a termination criterion for the global model is met:
\begin{enumerate*}
	\item Transmit the global model parameters $\vec{w}$ to the scheduled clients;
	\item Wait for the clients to run \cref{alg:Satellite SGD Procedure} and return their results;
	\item Aggregate the received gradients and update the global model parameters.
\end{enumerate*}
The difference between \cgls{fl} algorithms is in the computation of gradients in \cref{alg:Satellite SGD Procedure} and the update rule in Step 3). While we focus on FedAvg here, the extension to many other algorithms is trivial.

Consider global iteration $n$ and assume all clients are scheduled to participate in this iteration. Plain FedAvg implements the update rule
$\vec w^{n+1} = \frac{1}{D} \sum_{k = 1}^K D_k \vec w_k^{n}$. An equivalent update rule based on effective gradients $\vec g_k(\vec w^n_k)$, as {\color{black}represented in} \cref{alg:Satellite SGD Procedure}, is
\begin{equation} \label{eq: FedISLSyncc}
\begin{aligned}
\vec w^{n+1} &= \frac{1}{D} \sum_{k=1}^K D_k \left( \vec w_k^{\color{black}n} - \vec w_k^{n,0} \right) + \frac{1}{D} \sum_{k=1}^K D_k \vec w_k^{n,0}
	\\&= \vec w^{n} + \frac{1}{D} \sum_{k=1}^K D_k \vec g_k(\vec w_k^n).
 \end{aligned}
\end{equation}
A common generalization is to add a global server learning rate $\eta_s$ to this update rule, i.e.,
$\vec w^{n+1} = \vec w^{n} - \eta_s \vec\gamma^n$, where $\vec\gamma^n = -\frac{1}{D} \sum_{k=1}^K D_k \vec g_k(\vec w_k^n)$ \cite{wang2021field}.

The complete algorithm for \cgls{ps} operation in the \cgls{sfl} scenario, in relation to the client update procedure in \cref{alg:Satellite SGD Procedure}, is given in \cref{alg:syncPS}. This is a modified version of the delay-tolerant FedAvg implementation in \cite{WCL_fedsat}. It is initialized in line~\ref{alg:syncPS:init}, where a set of client clusters $\mathscr C$ is defined. This is required for the efficient use of \cglspl{isl} in \cref{sec:routing}. The idea is to treat each cluster as if it were an individual user, receiving the current parameter vector only once and also delivering a single effective gradient per global iteration. For now, as well as for scenarios without \cglspl{isl}, it can be assumed that satellite/client $k$ is mapped to cluster $\mathcal C_k$, with $P$ clusters in total. The \cgls{ps} maintains the training process until the sequence $\{ \vec w^1, \vec w^2, \dots \}$ satisfies the termination criterion in line~\ref{alg:syncPS:termination}. Each iteration of this outer loop corresponds to a global iteration, counted as $n$. The sets $\mathcal T^n$ and $\mathcal R^n$ track the transmission of $\vec w^n$ and reception of the gradient update to $\vec w^n$ per client group, respectively. Hence, the inner loop in line~\ref{alg:syncPS:inner}--\ref{alg:syncPS:innerend} runs until updates have been received from all clusters in $\mathscr C$. This loop blocks until a satellite connects to the \cgls{ps} in line~\ref{alg:syncPS:wait}. Note that this could also mean continuing a connection that was not terminated the previous iteration. A message from the satellite is received that either requests the transmission of the current parameter vector or contains the update from the satellite's cluster. 

\begin{algorithm}
\caption{Synchronous PS Operation}\label{alg:syncPS}
\small
\begin{algorithmic}[1]
	\State \textbf{initialize} global iteration $n = 0$, model $\vec{w}^0$,\\\hspace{2em} and client clusters $\mathscr C = \{ \mathcal C_1, \mathcal C_2, \dots \}$ \label{alg:syncPS:init}
	\Statex
	\While {termination criterion not met} \label{alg:syncPS:termination}
		\State Set $n \gets n + 1$, $\mathcal T^n = \mathcal R^n = \emptyset$, $\vec w^n \gets \vec w^{n-1}$ \label{alg:syncPS:innerinit}
		\Statex
		\While {$|\mathcal R^n| < |\mathscr C|$} \label{alg:syncPS:inner}
			\State Wait until connection from satellite $k$: \label{alg:syncPS:wait}
			\State Receive message $m$
			\State Find $p$ such that $k \in \mathcal C_p$
			\Statex
			\If {$p \notin \mathcal T^n$ \textbf{and} $m$ is request for data}
				\State Transmit $\vec w^{n-1}$ to satellite $k$ \label{alg:syncPS:transmit}
				\State Upon successful transfer, add $p$ to $\mathcal T^n$ \label{alg:syncPS:transmit:succ}
			\parElsIf {$p \notin \mathcal R^n$ \\\textbf{and} $m$ contains gradient update $\vec\gamma_p$}
				\State $\tilde{\vec\gamma}_p \gets \Call{UncompressGradient}{\vec\gamma_p}$ \label{alg:syncPS:uncompress}
				\State $\vec w^n \gets \vec w^n + \frac{1}{D} \tilde{\vec\gamma}_p$ \label{alg:syncPS:aggregate}
				\State Add $p$ to $\mathcal R^n$
				\State Acknowledge reception to $k$  \label{alg:syncPS:aggregate:ack}
			\EndIf
			\Statex
			\If {$|\mathcal R^n| < |\mathscr C|$}
				\State Terminate connection \label{alg:syncPS:disconnect}
			\EndIf
		\EndWhile \label{alg:syncPS:innerend}
	\EndWhile
\end{algorithmic}
\end{algorithm}

If the satellite requests transmission of $\vec w^n$ and it was not yet delivered to its cluster, it will be transmitted in line~\ref{alg:syncPS:transmit}. Successful reception must be acknowledged by the satellite. A possible implementation is the Bundle protocol's custody transfer~\cite{rfc5050,rfc9171}. Then, this cluster is marked as having received the transmission in line~\ref{alg:syncPS:transmit:succ}. All subsequent transmission requests for $\vec w^n$ to satellites of this cluster will be rejected by terminating the connection in line~\ref{alg:syncPS:disconnect}. If the satellite transmits an update $\vec\gamma_p$ to $\vec w^n$ and the cluster has not yet transmitted an update in the iteration, the compression applied in line~\ref{alg:ssgd:compress} of \cref{alg:Satellite SGD Procedure} is decoded in line~\ref{alg:syncPS:uncompress}. Without compression, \textsc{UncompressGradient} is the identity function. Then, the update rule \cref{eq: FedISLSyncc} is applied in line~\ref{alg:syncPS:aggregate}. incrementally (see also line~\ref{alg:syncPS:innerinit}) and relies on the received gradient already being scaled by $D_k$. Unless the current iteration is finished, the connection is terminated in \cref{alg:syncPS:disconnect}. 

\begin{figure*}
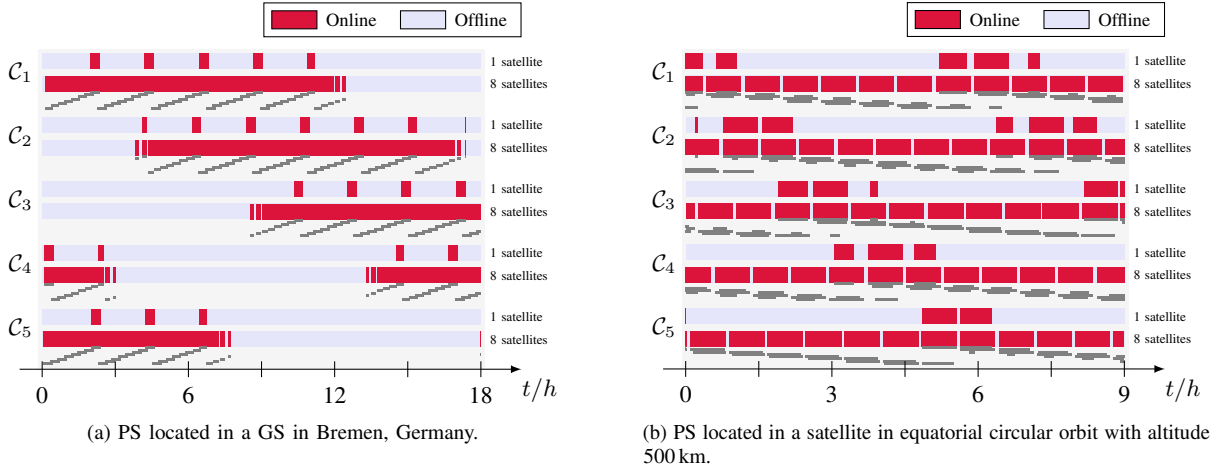

	\centering%
	\subfloat[\cGls{ps} located in a \cgls{gs} in Bremen, Germany.]{%
		\input{images/connectivity_sporadic.tikz}%
		\label{fig:pattern:sporadic}%
	}%
	\hfil%
	\subfloat[\cGls{ps} located in a satellite in {\color{black}equatorial} circular orbit with altitude \SI{500}{\km}.]{%
		\input{images/connectivity_dense.tikz}%
		\label{fig:pattern:dense}%
	}%
	\caption{Connectivity towards the \cgls{ps} from within a \SI{60}{\degree}: 40/5/1 Walker delta constellation. That is, a constellation of 40 satellites having \SI{60}{\degree} inclined circular orbits and altitude \SI{2000}{\km}. The satellites are distributed evenly among five orbital planes, which are spaced equidistantly around Earth. Clusters $\mathcal C_p$ are defined as either a single satellite per orbital plane or all satellites within an orbital plane. In the second case, a cluster is considered having a connection to the \cgls{ps} if at least one satellite of the cluster can communicate with the \cgls{ps}. Per-satellite connectivity towards the \cgls{ps} is displayed in gray below the cluster connectivity.}
	\label{fig:pattern}
\end{figure*}

\subsection{Asynchronous Orchestration}
As opposed to synchronous operation, in asynchronous \cgls{ps} operation{\color{black},} the \cgls{ps} does not delay the global model update until all clients have delivered their local updates. Instead, it opportunistically  incorporates received gradient updates into a new iteration of the global model, which is subsequently distributed to the clients. Consequently, clients simultaneously operate on different version of the model parameters and gradient updates are typically based on an outdated version of the model parameters, reducing the convergence speed and, potentially, numerical problems. For \cgls{sfl}, this \emph{stalenenss}, i.e., the age of the local model with respect to the current global model, is bounded due to the quasiperiodicity of this scenario. Hence, we are operating in the partially asynchronous domain, which leads to, generally speaking, much better convergence properties than in totally asynchronous scenarios with unbounded delays \cite{Bertsekas2015}.

Inspired by \cite{WCL_fedsat}, an asynchronous version of \cref{alg:syncPS} is proposed in \cref{alg:asyncPS}. It consists of the same functional blocks  as \cref{alg:syncPS}: Wait for incoming connections in lines~\ref{alg:asyncPS:wait}--\ref{alg:asyncPS:wait2}, incorporate and acknowledge gradient updates in lines~\ref{alg:asyncPS:uncompress}--\ref{alg:asyncPS:aggregate:ack}, run until global convergence (lines~\ref{alg:asyncPS:termination}--\ref{alg:asyncPS:termination2}), and transmit the current version of the model parameters in lines~\ref{alg:asyncPS:transmit}--\ref{alg:asyncPS:transmit:succ}. The main difference is that, in \cref{alg:syncPS}, the nested loop ensures that every cluster adds their update to the global model before this new version is transmitted to any client. Instead, \cref{alg:asyncPS} directly incorporates every received update and immediately starts using this new version to answer requests for data. To avoid race conditions within the clusters, \cref{alg:asyncPS} uses the set $\mathcal A$ to track active clusters, i.e., clusters that received the model parameters and did not yet return an update.
Another difference is that the global termination criterion, after being first met, might not remain valid after receiving the outstanding updates from active clusters. To handle this, whenever a new global model satisfies the termination criterion, all inactive clusters in $\mathcal B$, including the currently connected, are blocked from further computations. This is repeated until no active cluster remains. If the termination criterion is violated for any update, all blocks are removed in line~\ref{alg:asyncPS:block} and the algorithm resumes normal operation.

\begin{algorithm}
\caption{Asynchronous PS Operation}\label{alg:asyncPS}
\begin{algorithmic}[1]
	\State \textbf{initialize} global iteration $n = 0$, model $\vec{w}^0$, $\mathcal A = \mathcal B = \emptyset$, \\\hspace{2em} and client clusters $\mathscr C = \{ \mathcal C_1, \mathcal C_2, \dots \}$ \label{alg:asyncPS:init}
	\Loop
		\State Wait until connection from satellite $k$: \label{alg:asyncPS:wait}
		\State Receive message $m$
		\State Find $p$ such that $k \in \mathcal C_p$ \label{alg:asyncPS:wait2}
		\If {$p \in \mathcal A$ and $m$ contains gradient update $\vec\gamma_p$}
			\State $\tilde{\vec\gamma}_p \gets \Call{UncompressGradient}{\vec\gamma_p}$ \label{alg:asyncPS:uncompress}
			\State $\vec w^n \gets \vec w^n + \frac{1}{D} \tilde{\vec\gamma}_p$ \label{alg:asyncPS:aggregate}
			\State $\mathcal A \gets \mathcal A \setminus \{ p \}$
			\State Acknowledge reception to $k$ \label{alg:asyncPS:aggregate:ack}
			\State Wait for new message $m$

			\Statex
			\If {termination criterion is met} \label{alg:asyncPS:termination}
				\State $\mathcal B \gets \{ 1, 2, \dots, P \} \setminus \mathcal A$
				\If {$\mathcal A = \emptyset$}
					\State Terminate loop (and connection)
				\EndIf
			\Else
				\State $\mathcal B \gets \emptyset$ \label{alg:asyncPS:block}
			\EndIf
		\EndIf\label{alg:asyncPS:termination2}
		\If {$p \notin \mathcal A \cup \mathcal B$ and $m$ is request for data}
			\State Transmit $\vec w^{n-1}$ to satellite $k$ \label{alg:asyncPS:transmit}
			\State Upon successful transfer, add $p$ to $\mathcal A$ \label{alg:asyncPS:transmit:succ}
		\EndIf
		\State Terminate connection \label{alg:asyncPS:disconnect}
	\EndLoop
\end{algorithmic}
\end{algorithm}

\section{Intra-Orbit Aggregation for Satellite FL} \label{sec:routing}
Convergence in \cgls{sfl} is mainly impaired by the connectivity bottleneck between satellites and the \cgls{ps}.
We have discussed algorithmic approaches to this obstacle in the previous section. With the availability of \cglspl{isl}, these can be complemented by multi-hop routing to close gaps in connectivity. \Cref{fig:pattern} illustrates the potential gain for a constellation having five orbital planes, each with eight satellites, displaying the connection of a single satellite from each {\color{black}orbital} plane towards a \cgls{ps}. Using multi-hop routes within the orbital plane leads to significantly prolonged online periods as compared to the sporadic point-to-point connectivity. For simplicity, we focus on synchronous orchestration first and review the specifics for asynchronous orchestration in \cref{sec:routing:async}.

Communication for \gls{fl} involves two primary tasks, parameter vector distribution and collection of gradient updates. A direct approach to implementing these, leveraging \cglspl{isl} for multi-hop routing, are conventional \cgls{dtn} techniques for satellite networks, e.g., \cgls{cgr} in combination with the Bundle protocol \cite{rfc4838,Araniti2015,Fraire2021}. Then, parameter vector distribution is a multicast message from the \cgls{ps} towards all satellites in $\mathcal K$, while gradient updates are communicated as unicast messages from individual satellites towards the \cgls{ps}. However, the number of these unicast transmissions scales quadratically in the number of satellites per orbital plane. This can be reduced to a linear increase with in-network aggregation.

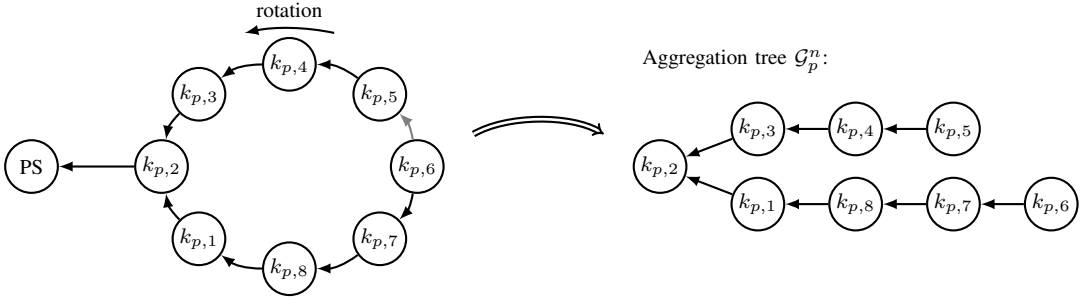
\begin{figure*}
	\centering
	\begin{tikzpicture}[grow=right, every node/.style={font=\footnotesize}, sat/.style={circle,draw=black, inner sep = 0, minimum width = .7cm, color = black}, edge from parent/.style={draw,latex-,color = black},sibling distance=10mm,level distance=13mm, every path/.append style={thick}]

		\begin{scope}[yscale = .8]
			\foreach \i in {1, 2, ..., 8}
				\node [sat] (sat\i) at (-\i*45-90:1.7cm) {$k_{p,\i}$};

			\draw[-latex,radius=2.3cm] (75:2.3) arc [start angle=75, delta angle=30] node [midway, yshift=7pt] {rotation} ;
		\end{scope}

		\foreach \i [remember=\i as \lasti (initially 2)] in {1, 8, 7, 6}
			\draw[out=-18,in=198,relative,latex-] (sat\lasti) to (sat\i);

		\foreach \i [remember=\i as \lasti (initially 2)] in {3, 4, 5}
			\draw[out=-18,in=198,relative,-latex] (sat\i) to (sat\lasti);

			\draw[out=-18,in=198,relative,-latex, color = gray] (sat6) to (sat5);

		\node [sat] (ps) [left=of sat2] {PS};
		\draw [latex-] (ps) -- (sat2);

		\begin{scope}[every node/.append style={sat}]
			\node [right=2.5cm of sat6] (root) {$k_{p,2}$}
				child {node {$k_{p,1}$} child {node {$k_{p,8}$} child {node{$k_{p,7}$} child {node{$k_{p,6}$}}}}}
				child {node {$k_{p,3}$} child {node {$k_{p,4}$} child {node{$k_{p,5}$}}}};
		\end{scope}
		\node [anchor=west] at ($(root.west)+(0,1.4cm)$) {Aggregation tree $\mathcal G_p^n$:};

		\draw [shorten <= 5mm, shorten >= 5mm, thick, double distance=1pt, -{Implies[]}, bend left] (sat6) to (root);
	\end{tikzpicture}
	\caption{Routing tree for incremental aggregation in orbital plane $p$. Satellite $k_{p,2}$ acts as sink node. Satellite $k_{p,6}$ has two shortest-path routes to sink. This is resolved unambiguously by the routing algorithm.}
	\label{fig: Routing tree for aggregation phase}
\end{figure*}

\subsection{Incremental Aggregation} \label{sec:routing:agg}
Consider the \cgls{ps} update rule from \cref{eq: FedISLSyncc}, within the global iteration $n$. The \cgls{ps} is primarily interested in the sum of effective gradients instead of individual gradient updates. Hence, the updates from all satellites within an orbital plane $\mathcal K_p$ can be collected at a single satellite $s_p^n \in \mathcal K_p$ for transmission to the \cgls{ps}. Instead of transmitting the individual gradient updates $\{ D_k \bar{\vec g}_k(\vec w_k^n) \}_{k\in\mathcal K_p}$, this satellite sends the linear combination $\sum_{k\in\mathcal K_p} D_k \bar{\vec g}_k(\vec w_k^n)$ to the \cgls{ps}. Following \cref{sec:ps}, the \cgls{ps} then treats the satellites in $\mathcal K_p$ as the client cluster $\mathcal C_p$ and incorporates their joint update.

This approach can be extended to intra-cluster communication. Suppose the sink satellite $s^n_p$ is known to all satellites in $\mathcal K_p$. Then, each satellite can easily determine a shortest-path in-tree, rooted at and oriented towards $s^n_p$, for $\mathcal K_p$ \cite[\S 9.6]{Deo1974}. For $K_p$ odd, this tree is unique. For $K_p$ even, two equal length paths exist for the satellite farthest away from $s^n_p$. Denote this satellite $k_{p,i}$ and resolve this ambiguity by choosing the tree where $k_{p,i+1}$ is the parent node of $k_{p,i}$.
We denote this directed graph as the \emph{aggregation tree} $\mathcal G_p^n = (\mathcal K_p, \mathcal E_p^n)$, with vertex set $\mathcal K_p$ and edge set $\mathcal E_p^n$. Its computation is illustrated in \Cref{fig: Routing tree for aggregation phase}.

Let $\mathcal N_{\mathcal G_p^n}^-(k) = \{ y \in\mathcal K_p : (y, k) \in \mathcal E_p^n \}$ be the incoming neighborhood of $k$ on $\mathcal G_p^n$. 
Based on $\mathcal G_p^n$, each satellite $k\in\mathcal K_p$ knows exactly which gradient updates it needs to forward. After local computation completes, node $k$ waits until all updates from the satellites in $\mathcal N_{\mathcal G_p^n}^-(k)$ are received via \cglspl{isl}. This might have already happened during the local computation phase. Then, the outgoing partial aggregate of satellite $k$ is computed as
\begin{equation} \label{eq:partialaggregate}
	\vec\gamma^n_k = {D_k} \bar{\vec{g}}_k(\vec{w}_k^{n}) + \sum_{y \in \mathcal N_{\mathcal G_p^n}^-(k)} \vec\gamma^n_y
\end{equation}
and transmitted via \cgls{isl} to the next hop $p \in \mathcal N_{\mathcal G_p^n}^+(k) = \{ y \in\mathcal K_p : (k, y) \in \mathcal E_p^n \}$. If $k$ is the current\footnote{Please refer to \cref{sec:routing:failure} for a failure handling procedure that might change the sink node after initial assignment.} sink node, i.e., $|\mathcal N_{\mathcal G_p^n}^+(k)|= 0$, $\vec\gamma^n_k$ is the joint aggregate of client cluster $\mathcal C_g = \mathcal K_p$ and transmitted to the \cgls{ps} once the link becomes available.

\subsection{Parameter Vector Distribution} \label{sec:routing:distr}
\Cref{alg:syncPS,alg:asyncPS} treat client clusters as if they were a single node, i.e., the \cgls{ps} expects to receive a single (joint) update per client cluster and transmits the model parameter vector only once per global iteration to each cluster. The previous subsection provides ample reason for the first design choice. Transmitting the parameter vector only once is also sensible for several reasons. First of all, it is what ideally happens if the \cgls{ps} transmits a multicast message towards the nodes in $\mathcal K_p$ using the Bundle protocol. As noted before, using the Bundle protocol's custody transfer (or similar) to safeguard point-to-point transmissions involving the \cgls{ps} is a fundamental assumption in this paper. Second, using multicast messages and multi-hop routing considerably reduces delay and the communication effort for the \cgls{ps}. Finally, this provides a natural leader election mechanism for coordinating the sink node selection.

The \cgls{ps} sends the model parameters $\vec w^n$ in iteration $n$ to a single satellite $k$ per cluster $\mathcal C_p$. Satellite $k$ is then responsible for determining an appropriate sink node $s_p^n$ using the procedure described in \cref{sec:routing:pred}. Subsequently, satellite $k$ propagates its routing decision together with $\vec w^n$ through both \cglspl{isl} to its neighbors $\mathcal N(k)$ and starts the local training $\textsc{ClientOpt}(\vec w^n)$.
All other satellites, upon reception of new model parameters $\vec w^n$ via one of the \cglspl{isl}, forward the model parameters to their next neighbor that did not receive $\vec w^n$ and start the local training $\textsc{ClientOpt}(\vec w^n)$.
Subsequent receptions of $\vec w^n$ are silently dropped upon reception. This ensures that the propagation of $\vec w^n$ through the orbital plane stops once every satellite received it. 

\subsection{Predictive Routing} \label{sec:routing:pred}
The final missing piece to this routing approach is a procedure to determine the sink node $s^n_p$. This decision must be made before the local training procedure completes on any satellite in the client cluster.
From a timing perspective, the ideal sink node completes computing \cref{eq:partialaggregate} immediately before its link towards the \cgls{ps} becomes available.
Predicting the future state of a satellite constellation is possible with high accuracy due to the determinism of satellite movement \cite{Vetter2007,Vallado2008,Vallado2013}. Here, we assume the availability of an orbital propagator with low computational complexity and (relatively) low precision like SGP4 \cite{SGP4,Vallado2008}. We also assume availability of fairly recent information on the constellation state, e.g., in the form of \glspl{tle} \cite{TLE,SGP4}, distributed by the satellite operator. The sink node can be determined from the orbital positions of the \cgls{ps} and satellites $\mathcal K_p$ at the time when intra-orbit aggregation completes. This requires a prediction of the time $T_p^n$ to 1) distribute the model parameters; 2) compute the local updates; and 3) deliver these updates to candidate sink nodes. The computation time can be estimated using $t_l(k)$ in \cref{eq: Time of learning}. The time to transmit a vector $\vec v$ via intra-orbit \cgls{isl} between two satellites $k_1$ and $k_2\in\mathcal N(k_1)$ is upper bounded as
$t_c(\vec v, t; k_1, k_2) \le \frac{S(\vec v)}{\rho_{k_1,k_2}} + \frac{\max_t d(t; k_1, k_2)}{c_0}$,
where $c_0$ is the vacuum speed of light, $S(\vec v)$ is the storage size of $\vec v$ and $d(t; k_1, k_2)$ is constant in $t$ for circular orbits. For uncompressed gradients, $S(\vec w^n) = n_d \omega$, where $\omega$ is the storage size of a single element of model parameter, usually \SI{16}-{bit} or \SI{32}-{bit} floating point number.

A precise solution for $T_p^n$ should take into account that satellites start their learning procedure at different times depending on the number of hops the global model parameters need to travel. Even when distribution takes place over shortest-distance paths, the completion time is different for each potential sink node in $\mathcal K_p$. Hence, predictions on the constellation state must be made for up to $K_p$ time instants and the computational complexity for determining the sink node  scales at least linearly in $K_p$.
A considerably simpler approach is to assume distribution takes place over shortest-distance paths and make a worst-case estimate for the aggregation phase. Combined with assuming a symmetric orbital distribution of satellites and equal \gls{isl} capabilities, i.e., $d_p = \max_t d(t; k_1, k_2)$ and $\rho_p = \rho_{k_1,k_2}$ for all $k_1 \in\mathcal K_p, k_2 \in\mathcal N(k_1)$, we obtain
\begin{equation}
\begin{aligned}
	T_{p}^n &\le
	\left\lceil \frac{K_p}{2} \right\rceil \left( \frac{S(\vec w^n)}{\rho_p} + \frac{d_p}{c_0} \right)
	+
	\max_{k\in\mathcal K_p} t_l(k)\\
	&+
	\left\lceil \frac{K_p}{2} \right\rceil \left( \frac{\max_{k\in\mathcal K_p}S(\bar{\vec g}_k(\vec w_k^n))}{\rho_p} + \frac{d_p}{c_0} \right).
 \end{aligned}
\end{equation}
Under constant-length gradient compression and equal in-cluster computation times, becomes:
\begin{equation}\label{eq: aggregationtime}
	T_{p}^n \le \hat T_p^n = 
	t_l(k_1)
	+
	\left\lceil \frac{K_p}{2} \right\rceil \left(
		\frac{S(\vec w^n) + S(\bar{\vec g}_{k_1}(\vec w_{k_1}^n))}{\rho_p}
		+ \frac{2 d_p}{c_0}
	\right)
\end{equation}
for any $k_1 \in\mathcal K_p$. This bound overestimates the time for communications. However, this part is likely small compared to the computation time. Taking into account the inaccuracies of orbital prediction,\footnote{\Gls{tle} data has an initial accuracy of roughly \SI{1}{km} and then decays quickly \cite{SGP4}.} local clock deviations, and computational delays due to, e.g., multitasking, interrupts, or priority scheduling, slightly overestimating the time to completion seems reasonable.
The major advantage of using the simple estimate in \cref{eq: aggregationtime} is that the constellation state needs only be predicted for a single time instant. Thus, the computational effort remains constant in $K_p$.

Based on the estimate $\hat T_p^n$, the satellite that took custody of $\vec w^n$ computes the positions of the \cgls{ps} and all satellites in its orbital plane at time $t_p^n + \hat T_p^n$, where $t_p^n$ is the expected local time after determining $s_p^n$. {\color{black}Among the satellites in communication range of the \cgls{ps}, the satellite with the longest remaining window for communication is selected as sink node $s_p^n$. Should this window be of insufficient length to transmit the parameter vector or if no satellite is in communications range of the \cgls{ps}, the next satellite of that orbital plane to contact the \cgls{ps} is selected as sink $s_p^n$. Then, the custodian satellite starts the model distribution process as described in \cref{sec:routing:distr}.}

\subsection{Failure Handling} \label{sec:routing:failure}
{\color{black}
	If the sink node completes the incremental aggregation after its communication window to the \cgls{ps} has closed, a critical routing error may occur. This can happen due to random factors in learning and transmission, such as multiprocessing loads and queuing delays. As a result, the sink may fail to deliver the aggregated parameters to the \cgls{ps} within the designated, initially planned time. 

To deal with these situations, we introduce failure handling. A straightforward scheme, termed {pass-to-neighbor}, could work as follows. If the sink satellite $s_p^n$ is unable to deliver the aggregated parameters to the \cgls{ps}, it transfers those parameters to a neighboring satellite. If the \cgls{ps} is visible, this neighbor sends them to the \cgls{ps}; otherwise, it relays them further to the next neighbor. This process of passing to the next neighbor, within a fixed direction, continues until a satellite can forward the parameters to the \cgls{ps}. 
Pass-to-neighbor may lengthen the whole \cgls{fl} process if
it takes an excessive time to find a satellite that can forward the aggregated parameters to the \cgls{ps}.

A more involved, yet practical, scheme is \textit{determine-new-sink}. Here, after finalizing the parameter aggregation, the sink satellite $s_p^n$ can designate a new sink $\hat{s}_p^{n}$, if the delay dictates that. In this regard, the sink $s_p^n$ uses visibility of satellites and the time required for \cgls{isl} transmissions 
to calculate the time it takes for the aggregated parameters to reach a potential new sink, chosen from the set of satellites in the orbit $p$, i.e. $k \in \mathcal K_p =\{1,2,\cdots, K_p \}$, as
\begin{equation} \label{eq:failure:time}
	t(k) =
	t_0 + 
	 h(k,s_p^n)  \left(
		\frac{S(\gamma_{s_p^n}^{n})}{\rho_p}
		+ \frac{d_p}{c_0} 
	\right)+ t_{g},
\end{equation}
where $t_0$ is the current time of $s_p^n$ and $h(k,s_p^n)$ is the number of hops between satellite $k$ and the sink $s_p^n$ in the orbit. The parameter $t_{g}$ is a guard time, introduced to mitigate other delays in the orbit that may occur due to the parameter transmission from $s_p^n$ to the newly designated sink $\hat{s}_p^{n}$. The satellite $k$ is chosen as the new sink $\hat{s}_p^{n}$ if it satisfies both the conditions of having the lowest $t(k)$, the time at which it can first visit the \cgls{ps}. Subsequently, the aggregated parameters along with the ID of new sink $\hat{s}_p^{n}$ are sent from $s_p^n$ to $\hat{s}_p^{n}$. This is done via a multi-hop connection over the \cglspl{isl} of satellites between $s_p^n$ and $\hat{s}_p^{n}$. The new sink satellite $\hat{s}_p^{n}$ transmits the parameters to the \cgls{ps} when the \cgls{ps} becomes visible.}

\subsection{Algorithm}
The satellite operation described above is summarized in \cref{alg:sat}. This procedure runs until a stop message is received from the \cgls{ps}, either directly or via \cgls{isl}. The parameter vector distribution is handled in lines~\ref{alg:sat:waitdistr}--\ref{alg:sat:waitdistr:end}. The training procedure is launched concurrently, e.g., in a separate thread, in line~\ref{alg:sat:learn}. \Cref{alg:sat} continues immediately with line~\ref{alg:sat:startagg}, which starts the incremental aggregation phase by computing $\mathcal G_p^n$. {\color{black}Lines~\ref{alg:sink_FH_start}--\ref{alg:sat:passsinkend} are for failure handling.}

If the satellite is the current sink node, it waits for incoming results and its own computation to finish.
Should the designated \cgls{ps} communication window pass in the meantime, {\color{black} the failure handling} is {\color{black}applied} according to the procedure described in \cref{sec:routing:failure}. It triggers the {\color{black}failure handling procedure in lines~\ref{alg:sink_FH_start}--\ref{alg:sat:passsinkend}}, which {\color{black}calculates the new sink}.
Otherwise, once all results are ready, the sink node computes the final cluster aggregate in line~\ref{alg:sat:clusteragg}, waits for the \cgls{ps} to become available, and transmits the results.

All other satellites execute lines~\ref{alg:sat:cagg1}--\ref{alg:sat:cagg2}. If the local training is complete and the expected incoming partial aggregate is received, the satellite computes \cref{eq:partialaggregate} and forwards the result to the next hop in line~\ref{alg:sat:incagg}. It then returns to line~\ref{alg:sat:waitdistr} to wait for a new global iteration or take over as sink node. If the satellite is tasked with taking over as {\color{black}new} sink node, {\color{black}it performs as line~\ref{alg:line:FH}.}

\begin{algorithm}
	\caption{Satellite Operation} \label{alg:sat}
	\noindent%
	\begin{minipage}[t]{0.5\textwidth-1pc}
	\begin{algorithmic}[1]
		\State \textbf{Initialize} global iteration $n = 0$, satellite ID $k = k_{p,i}$\\\hspace{2em} and orbital plane ID $p$
		\Statex

		\State \textbf{Wait for} incoming \cgls{isl} \textbf{or} start of PS connectivity window\label{alg:sat:waitdistr}%
		\If {received $(s_p^{n+1}, \vec w^{n+1})$ from satellite $l$}
			\State $n \gets n + 1$
			\State Forward $(s_p^n, \vec w^n)$ to $\mathcal N(k)\setminus \{ l \}$
		\ElsIf {connected to \cgls{ps}}
			\State Request parameters $\vec w^{n+1}$ and wait for reply
			\If {received $\vec w^{n+1}$}
				\State Acknowledge reception to \cgls{ps} and set $n \gets n+1$
				\State Compute $t_p^n + \hat T_p^n$ and determine $s_p^n$ \label{alg:sat:asyncmod1}
				\Statex\Comment cf.~\cref{sec:routing:pred}
				\State Transmit $(s_p^n, \vec w_p^n)$ to $\mathcal N(k)$
			\Else
				\State Goto line~\ref{alg:sat:waitdistr}
			\EndIf
		\EndIf \label{alg:sat:waitdistr:end}

		\Statex
		\State \textbf{Execute concurrently} $D_k \bar{\vec g_k}(\vec w_k^n) \gets \Call{ClientOpt}{\vec w^{n}}$\label{alg:sat:learn}
		\State Compute aggregation tree $\mathcal G_p^n$ \label{alg:sat:startagg}
		\Comment cf.~\cref{sec:routing:agg}
		\algstore{satop}
	\end{algorithmic}
	\end{minipage}%
	\hfill%
	\begin{minipage}[t]{0.5\textwidth-1pc}
	\begin{algorithmic}
		\algrestore{satop}
		\If {$k$ is sink satellite}
			\State Initialize {\color{black}failure handling} \label{alg:sat:skip}
			\Comment cf.~\cref{sec:routing:failure}
		\State \parbox[t]{\linewidth}{\hangindent=2em\textbf{Wait for} \textbf{\lbrack} \textsc{ClientOpt} \textbf{and} results from $\mathcal N^-_{\mathcal G_p^n}(k)$ \textbf{\rbrack{}}\\\textbf{until} \emph{\color{black}failure}\strut} \label{alg:sat:sinkwait}
			\If {failure}
     \State {\color{black}Compute $\vec\gamma_k^n$ as in \cref{eq:partialaggregate} \label{alg:sink_FH_start}}
						\State {\color{black} Calculate $\hat{s}_p^{n}$ based on \cref{eq:failure:time}\label{alg:sat:passsink}} 
						\State
                         {\color{black} {Transmit 
($\hat{s}_p^{n},\gamma_{s_p^n}^{n}$) to the next satellite}
\label{alg:sat:passsinkend}}
			\Else
				\State Compute $\vec\gamma_k^n$ as in \cref{eq:partialaggregate} \label{alg:sat:clusteragg}
				\State \textbf{Wait for} connection to \cgls{ps} \label{alg:sat:asyncmod2}
				\State Transmit $\vec\gamma_k^n$ and wait for acknowledgement
			\EndIf
		\Else \label{alg:sat:cagg1}
		\State \textbf{Wait for} \parbox[t]{.73\linewidth}{\textbf{\lbrack} \textsc{ClientOpt} \textbf{and} results from $\mathcal N^-_{\mathcal G_p^n}(k)$ \textbf{\rbrack{} or} {\color{black}($\hat{s}_p^{n},\gamma_{s_p^n}^{n}$)\strut}}
			\If {{\color{black}($\hat{s}_p^{n},\gamma_{s_p^n}^{n}$)}}		\label{alg:sat:waitdistr:sink}
              \If {$k$ is $\hat{s}_p^{n}$}
              \State {\color{black}{Goto line \ref{alg:sat:asyncmod2}}\label{alg:line:FH}}	   
             \Else
             \State {\color{black}{Transmit 
($\hat{s}_p^{n},\gamma_{s_p^n}^{n}$) to the next satellite}}        
             \EndIf
			\Else
				\State Compute $\vec\gamma_k^n$ as in \cref{eq:partialaggregate}
				\State Transmit $\vec\gamma_k^n$ to next hop $p \in \mathcal N^+_{\mathcal G_p^n}(k)$ \label{alg:sat:incagg}
			\EndIf
		\EndIf \label{alg:sat:cagg2}
		\State Goto line~\ref{alg:sat:waitdistr}
	\end{algorithmic}
	\end{minipage}
\end{algorithm}

\subsection{Asynchronous Orchestration} \label{sec:routing:async}
The primary reason for asynchronous orchestration is long connectivity gaps due to orbital mechanics. These outages are considerably reduced by the techniques developed in this section.
As there is no apparent benefit of per-client asynchronous updates over using (synchronous) client clusters, the system performance under asynchronous aggregation is expected to benefit significantly from incremental intra-orbit aggregation. In fact, the proposed \cgls{fl} system can be implemented such that the clients are agnostic to the \cgls{ps} operation.

However, in some scenarios the improved connectivity can result in a large number of updates from a small group of client clusters. This can lead to heavily biased solutions and other convergence issues. One such problematic scenario could be \cref{fig:pattern:sporadic} combined with a short learning time $t_l(k)$.
A simple solution to this issue is to set a maximum update frequency per cluster, i.e., fix a minimum time $T_u$ between updates. In a completely trusted system (as is likely the case in \cgls{sfl}), this can be implemented by
1) modifying line~\ref{alg:sat:asyncmod1} in \cref{alg:sat} to use $\max\{ \hat T_p^n,\ T_u \}$ instead of $\hat T_p^n$;
2) distributing the time stamp $t_p^n + T_u$ together with the model parameters and designated sink node; and
3) not sending any updates to the \cgls{ps} before $t_p^n + T_u$, unless this would result in missing the planned communication window (modify line~\ref{alg:sat:asyncmod2} in \cref{alg:sat}).
Alternatively, the \cgls{ps} could simply refuse sending the current global model parameters to a cluster if the update frequency is too high.

\section{Sparsification-based Gradient Compression} \label{sec:sparsification}
Synchronizing large-scale \cgls{ml} models requires transmitting massive amounts of data. A widely employed method to reduce the communication cost is to truncate the effective gradients prior to transmission to only contain the elements with largest magnitude, while setting the other elements to zero. This is known as \emph{sparsification} \cite{aji2017sparse,alistarh2018convergence}, as the resulting vectors are communicated in sparse vector encoding. With practical sparsification ratios $q$ in the range of \SIrange{1}{10}{\percent}, the size of a single effective gradient vector can be reduced by approximately \SIrange{80}{98}{\percent}.

However, while a single sparsified vector has a deterministic length of $\lfloor n_d q \rfloor$ nonzero elements, the sum of multiple sparse length-$\lfloor n_d q \rfloor$ vectors has nondeterministic length. This leads to variable transmission lengths and, hence, uncertainties in the predictive routing procedure. In this section, we develop an estimator for sparse vectors subject to incremental aggregation based on probabilistic modeling of the weight vectors. Before, we briefly review \cgls{fl} gradient sparsification.

\subsection{Gradient Sparsification for FL}
Each client transmits only the $\lfloor n_d q \rfloor$ largest magnitude elements from its effective gradient vector $\vec g_k(\vec w_k^n)$. This truncation operation is denoted as $\mathsf{Top}_{q}$. A common method to improve the training performance under gradient sparsification is to track the accumulated sparsification error in a residual vector $\vec\Delta_k^n$. The effect is that small magnitude elements, which would be ignored in every iteration, are aggregated over several iterations and will survive sparsification at some point. 

The complete sparsification procedure, for an effective gradient $\vec g_k(\vec w_k^n)$, is
to accumulate the previous sparsification error into the effective gradient as $\vec g_k^\mathrm{acc}(\vec w_k^n) \gets \vec g_k(\vec w_k^n) + \vec\Delta_k^{n-1}$, then compute the sparse gradient for transmission as $\bar{\vec g}_k(\vec w_k^n) \gets \mathsf{Top}_{q}(\vec g_k^\mathrm{acc}(\vec w_k^n))$, and update the sparsification error as $\vec\Delta_k^n \gets \vec g_k^\mathrm{acc}(\vec w_k^n) - \bar{\vec g}_k(\vec w_k^n)$.
Convergence of \cgls{sgd} with this sparsification procedure is established in \cite{alistarh2018convergence}. The \textsc{CompressGradient} procedure in \cref{alg:Satellite SGD Procedure} is implemented as exactly these three steps, with $\bar{\vec g}_k(\vec w_k^n)$ being the return value. Then, \textsc{UncompressGradient} simply converts the sparse vector back to its full-length representation, and the summation in \cref{eq:partialaggregate} is implemented as a conventional sparse vector addition \cite[\S 2]{Duff2017}.

\subsection{Predictive Routing for Sparse Incremental Aggregation} \label{Predictive Routing for Sparse Incremental Aggregation}

The predictive routing procedure in \cref{sec:routing:pred} relies on knowledge of the storage size $S(\bar{\vec g}_k(\vec w_k^n))$. With sparsification applied, the estimates in \cref{eq: aggregationtime,eq:failure:time} remain no longer valid, {\color{black}as the relevant length $S(\vec\gamma_k^n)$ is no} longer constant over the aggregation path. To this end, first consider the following lemma.
\begin{lemma} \label{lem:sparse}
	Consider $L$ \cgls{iid} random vectors $\vec X_1, \dots, \vec X_L$ of dimension $n_d$. Let $\tilde{\vec X}_l = \mathsf{Top}_q(\vec X_l)$ for all $l = 1, \dots, L$. Then, the expected number of nonzero elements in $\vec X_\Sigma = \sum_{l=1}^L \tilde{\vec X}_l$ is $n_d - n_d \left( 1 - \frac{n_a}{n_d} \right)^L$, where $n_a = \lfloor n_d q \rfloor$ is the number of nonzero elements in each summand. 
\end{lemma}
\begin{IEEEproof}
	Consider the vector $\vec X_l = (X_{l,1}, \dots, X_{l,n_d})$ and let $A_{l,i}$ be the event that $\vec X_{l,i}$ is zero after the $\mathsf{Top}_q$ operation. The probability that $A_{l,i}$ occurs is $1 - \frac{n_a}{n_d}$ \cite[Lemma 13.1]{Vaart1998}.
	Further, let $A_{i}$ be the event that element $i$ is zero after $\mathsf{Top}_q$ in all $L$ vectors. Then, due to independence,
	$\Pr(A_i) = \Pr\bigg(\bigcap_{l=1}^L A_{l,i}\bigg) = \prod_l \Pr( A_{l,i} ) = \bigg( 1 - \frac{n_a}{n_d} \bigg)^L\!.$
	Observe that $A_i$ is the event that the $i$th element of $\vec X_\Sigma$ is zero. Thus, the expected number of zero elements in $\vec X_\Sigma$ is
		$\mathds E\left[ \sum_{i=1}^{n_d} I(A_i) \right] = \sum_i \mathds E\left[ I(A_i) \right] = \sum_i \Pr(A_i) = n_d \bigg( 1 - \frac{n_a}{n_d} \bigg)^L\!,$
	where $I(\cdot)$ is the indicator function that takes value 1 if $A_i$ occurs and 0 otherwise.
\end{IEEEproof}

Based on this lemma, we can make a reasonable estimate on the number of transmitted bits during incremental aggregation.
\begin{proposition} \label{prop:sparse}
	The expected total number of transmitted bits over $H$ hops of incremental aggregation is upper bounded as
	$n_d (\omega + \lceil \log_2 n_d \rceil) \left[ H + 1 - \frac{n_d}{n_a} \left[ 1 - \left(1 - \frac{n_a}{n_d} \right)^{H+1} \right] \right]$,
	where $n_a = \lfloor n_d q \rfloor$.
\end{proposition}
\begin{IEEEproof}
	Let $P_h = (k_1, k_2, \dots, k_{h+1})$, for $h = 1, 2, \dots, H$, be an increasing sequence of nested paths. Consider incremental aggregation over path $P_H$ starting at $k_1$. Denote by $S(P_h)$ the total number of bits transmitted over path $P_h$. Then, $S(P_1) = S(\vec\gamma_{k_1}^n)$ and $S(P_h) = S(\vec \gamma_h^n) + S(P_{h-1})$ for $h > 1$, where $\vec\gamma_k^n$ is the outgoing aggregate at node $k$ as defined in \cref{eq:partialaggregate}. At $k_1$, the outgoing aggregate has storage size $S(\vec\gamma_{k_1}^n) = n_a (\omega + \lceil \log_2 n_d \rceil)$, where  $\lceil \log_2 n_d \rceil$ accounts for the storage space of element indices in the sparse representation \cite[\S 2]{Duff2017}.

	At subsequent nodes $k_h$, $h > 1$, the outgoing aggregate has size $S(\vec\gamma_{k_{h}}^n) = S(D_{k_{h}} \bar{\vec g}_{k_{h}}(\vec w_{k_{h}}^n) + \vec\gamma_{k_{h-1}}^n)$. Modelling $\vec w_k^n$ as an \cgls{iid} random vector and assuming the gradients of the nodes along $P_H$ are independent, we obtain 		$\mathds E[S(\vec\gamma_{k_{h}}^n)] = (\omega + \lceil \log_2 n_d \rceil) \left(n_d - n_d \left( 1 - \frac{n_a}{n_d} \right)^h \right)$ from \cref{lem:sparse}.
	Further, observe that $S(\vec\gamma_{k_1}^n) = n_d - n_d \left( 1 - \frac{n_a}{n_d} \right)$. Then, by recursion,
	$\mathds E[S(P_H)] = \sum_{h = 1}^H \mathds E[S(\vec\gamma_h^n)]= n_d (\omega + \lceil \log_2 n_d \rceil) \left[H - \sum_{h=1}^H\left( 1 - \frac{n_a}{n_d} \right)^h \right]$, and due to the geometric sum identity, $\mathds E[S(P_H)] = n_d (\omega + \lceil \log_2 n_d \rceil) \left[H + 1 - \frac{n_d}{n_a} \left(1 - \left( 1 - \frac{n_a}{n_d} \right)^{H+1}\right) \right]$.

	Finally, observe that for dependent gradients the positions of the nonzeros after $\mathsf{Top}_q$ will be correlated. Hence, with the notation from the proof of \cref{lem:sparse}, $\Pr(A_{l,i}) \le \Pr(A_{l,i} | A_{l,i-1} \cdots A_{l,1})$. Thus, $\Pr(A_i) \ge \left( 1 - \frac{n_a}{n_d} \right)^L$ and $\mathds E\left[ \sum_i I(A_i) \right] \ge n_d \left( 1 - \frac{n_a}{n_d} \right)^L$.
\end{IEEEproof}

Leveraging \cref{prop:sparse}, we replace $\left\lceil \frac{K_p}{2} \right\rceil S(\bar{\vec g}_{k_1}(\vec w_{k_1}^n)$ in \cref{eq: aggregationtime,eq:failure:time} with
\begin{equation} \label{eq:sparseeff}
	n_d (\omega \hspace{-0.08 em}+\hspace{-0.08 em}\lceil \log_2 n_d \rceil)\hspace{-0.4 em}\left[\hspace{-0.4 em} \left\lceil \frac{K_p}{2} \right\rceil\hspace{-0.2 em}+\hspace{-0.2 em}1\hspace{-0.2 em}- \frac{n_d}{n_a}\hspace{-0.2 em}\left[ 1\hspace{-0.2 em}-\hspace{-0.2 em} \left(1 - \frac{n_a}{n_d}\hspace{-0.2 em} \right)^{\left\lceil \frac{K_p}{2} \right\rceil+1} \right]\hspace{-0.2 em}\right],
\end{equation}
to account for sparsification in the predicted routing procedure. This will overestimate the communication effort as it does not account for the dependence between gradients.
{\color{black}As discussed in \cref{sec:routing:pred}, a certain amount of ``slack'' is inconsequential and might even improve overall performance due to reducing the probability of timing failures. However, should tighter bounds become necessary, training a simple \cgls{ml} estimator for the aggregated vector length, e.g., using reinforcement learning, appears sensible. Instead, tighter analytical bounds} would require assumptions on a random distribution for the elements of $\vec w$ and experimental calibration of correlation between gradients.

\section{Performance Evaluation} \label{sec: Performance Evaluation}
\def\Wdelta{\ensuremath{\text{W-}\Delta}\xspace}
\def\Wstar{\ensuremath{\text{W-}\star}\xspace}

We evaluate the performance of the proposed system design for four representative scenarios. The worker satellites are organized either in a \SI{60}{\degree}: 40/5/1 Walker {\color{black}d}elta or a \SI{85}{\degree}: 40/5/1 Walker star constellation, both at an altitude of \SI{2000}{\km}. The notation $i\!:\!t/p/f$ indicates a constellation with $p$ evenly spaced circular orbital planes at inclination $i$, each having $t/p$ equidistant satellites. The phasing parameter $f$ defines the relative shift in \cgls{raan} between adjacent orbital planes and amounts to \SI{9}{\degree} in this particular setup \cite{leyva2022ngso}. Subsequently, we identify these constellations as \Wdelta and \Wstar, respectively.
These constellations are combined with a \cgls{ps} located either in a terrestrial \cgls{gs} located in Bremen, Germany, or in a \cgls{leo} satellite orbiting at an altitude of \SI{500}{\km} in the equatorial plane. 
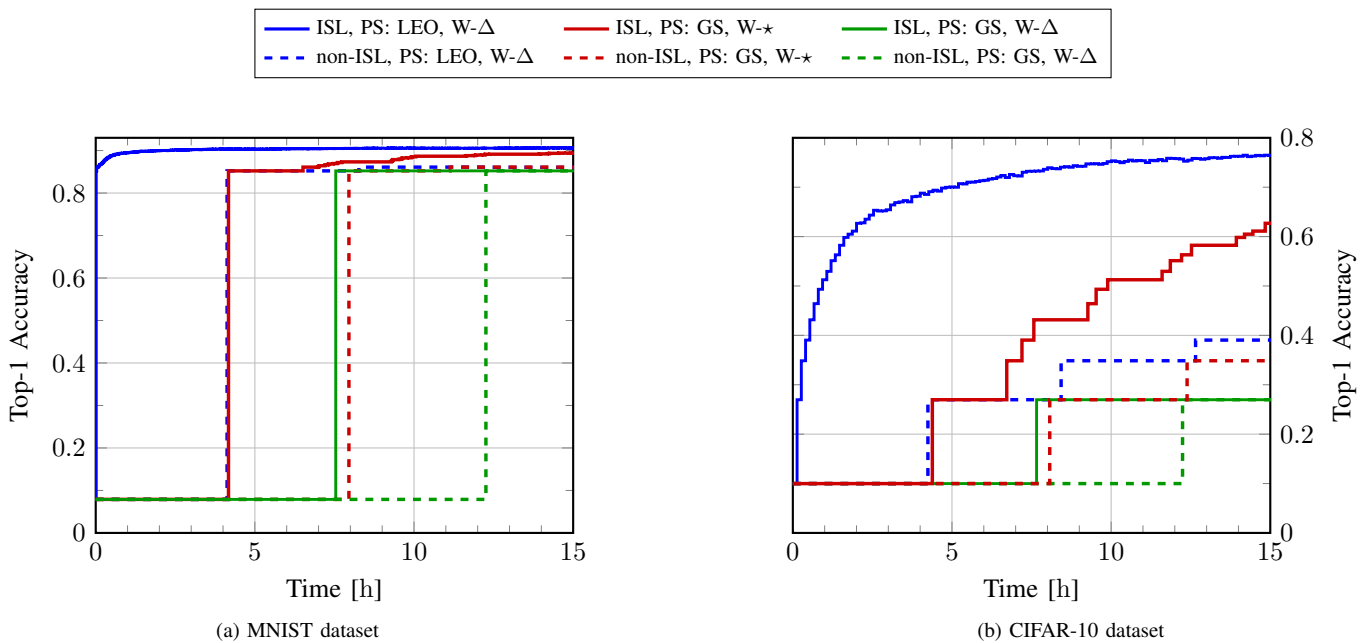
\begin{figure*}
	\centering
	\ref{fig:sync_mnist_cifar10_data:legend}\\[0.6em]
	\subfloat[MNIST dataset]{%
		\begin{tikzpicture}
			\begin{axis} [
				thick,
				xlabel={Time [$\si{\hour}$]},
				ylabel={Top-1 Accuracy},
				ylabel near ticks,
				grid=major,
				minor x tick num = 4,
				minor y tick num = 1,
				xmin = 0,
				ymin = 0,
				ymax = .93,
				xmax = 15,
				xtick = {0,5,...,15},
				legend pos=south east,
				legend cell align=left,
				cycle list name=default,
				legend image post style={very thick},
				no markers,
				width=.94*\axisdefaultwidth,
				legend entries = {\footnotesize{\cgls{isl}\text{,} PS: LEO\text{,} \Wdelta \hspace{1 cm}}, \footnotesize{non-\cgls{isl}\text{,} PS: LEO\text{,} \Wdelta \hspace{1 cm}}, \footnotesize{\cgls{isl}\text{,} PS: GS\text{,} \Wstar \hspace{1 cm}}, \footnotesize{non-\cgls{isl}\text{,} PS: GS\text{,} \Wstar \hspace{1 cm}}, \footnotesize{\cgls{isl}\text{,} PS: GS\text{,} \Wdelta \hspace{1 cm}},   \footnotesize{non-\cgls{isl}\text{,} PS: GS\text{,} \Wdelta} },
				legend to name = {fig:sync_mnist_cifar10_data:legend},
				legend columns = 2,
				transpose legend,
				legend style={/tikz/every even column/.append style={column sep = 2mm}},
			] 

				\addplot[color=blue,line width=1.1pt] table [x=Time_FedISL_Sync_MEO, y=Acc_FedISL_Sync_MEO, col sep=comma] {images/FedISL_FedNonISL_MNIST_NonIID.csv};

\addplot[color=blue,line width=1.3pt, dashed] table [x=Time_FedNonISL_Sync_MEO, y=Acc_FedNonISL_Sync_MEO, col sep=comma] {images/FedISL_FedNonISL_MNIST_NonIID.csv};

\addplot[color=black!20!red,line width=1.3pt] table [x=Time_FedISL_Sync_Bremen_walker_star, y=Acc_FedISL_Sync_Bremen_walker_star, col sep=comma] {images/FedISL_FedNonISL_MNIST_NonIID_walker_star.csv};

\addplot[color=black!20!red,line width=1.3pt, dashed] table [x=Time_FedNonISL_Sync_Bremen_walker_star, y=Acc_FedNonISL_Sync_Bremen_walker_star, col sep=comma] {images/FedISL_FedNonISL_MNIST_NonIID_walker_star.csv};

\addplot[color=black!40!green,line width=1.1pt] table [x=Time_FedISL_Sync_Bremen, y=Acc_FedISL_Sync_Bremen, col sep=comma] {images/FedISL_FedNonISL_MNIST_NonIID.csv};

\addplot[color=black!40!green,line width=1.3pt, dashed] table [x=Time_FedNonISL_Sync_Bremen, y=Acc_FedNonISL_Sync_Bremen, col sep=comma] {images/FedISL_FedNonISL_MNIST_NonIID.csv};

			\end{axis}
		\end{tikzpicture}%
		\label{fig:Sync_MNIST}%
	}
	\hfill%
	\subfloat[CIFAR-10 dataset]{%
		\begin{tikzpicture}
			\begin{axis} [
				thick,
				xlabel={Time [$\si{\hour}$]},
				ylabel={Top-1 Accuracy},
				ylabel near ticks,
				grid=major,
				minor x tick num = 4,
				minor y tick num = 1,
				xmin = 0,
				ymin = 0,
				ymax = .8,
				xmax = 15,
				xtick = {0,5,...,15},
				legend pos=south east,
				legend cell align=left,
				cycle list name=default,
				legend image post style={very thick},
				no markers,
				width=.94*\axisdefaultwidth,
				legend entries = {},
				yticklabel pos = right,
			]

			\addplot[color=blue,line width=1.1pt] table [x=Time_FedISL_Sync_MEO, y=Acc_FedISL_Sync_MEO, col sep=comma] {images/FedISL_FedNonISL_CIFAR.csv};

\addplot[color=blue,line width=1.3pt, dashed] table [x=Time_FedNonISL_Sync_MEO, y=Acc_FedNonISL_Sync_MEO, col sep=comma]{images/FedISL_FedNonISL_CIFAR.csv};

\addplot[color=black!40!green,line width=1.1pt] table [x=Time_FedISL_Sync_Bremen, y=Acc_FedISL_Sync_Bremen, col sep=comma]{images/FedISL_FedNonISL_CIFAR.csv};

\addplot[color=black!40!green,line width=1.3pt, dashed] table [x=Time_FedNonISL_Sync_Bremen, y=Acc_FedNonISL_Sync_Bremen, col sep=comma]{images/FedISL_FedNonISL_CIFAR.csv};

\addplot[color=black!20!red,line width=1.3pt] table [x=Time_FedISL_Sync_Bremen_star, y=Acc_FedISL_Sync_Bremen_star, col sep=comma] {images/FedISL_FedNonISL_CIFAR_star.csv};

\addplot[color=black!20!red,line width=1.3pt, dashed] table [x=Time_FedNonISL_Sync_Bremen_star, y=Acc_FedNonISL_Sync_Bremen_star, col sep=comma] {images/FedISL_FedNonISL_CIFAR_star.csv};

			\end{axis}
		\end{tikzpicture}%
		\label{fig:Sync_CIFAR10}%
	}%
	\caption{Test accuracy with respect to wall-clock time. Synchronous orchestration for the MNIST and CIFAR-10 datasets with non-\cgls{iid} distributions, considering both terrestrial and non-terrestrial PS. i.e., the GS in Bremen and an LEO satellite. Note that \cgls{isl} and non-\cgls{isl} stand for the FedAvg with and without \cgls{isl} algorithms respectively.}
 \label{fig:Sync_MNIST_CIFAR10}
\end{figure*}

Communication links are modelled as complex Gaussian channels with \cgls{fspl}. Then, the maximum achievable rate between nods $k$ and $i$ using a bandwidth $B$ is $\rho(k,i) = B \log_2 \left(1+\mathrm{SNR}(k,i)\right)$. The 
$\mathrm {SNR}(k,i) = \frac{P_t G_{k}(i) G_{i}(k) }{N_0  L(k,i)}$ is defined by the transmit power $P_t$, the noise spectral density $N_0 = k_B T B$ at receiver temperature $T$, and average antenna gains $G_j(l)$ at node $j$ towards node $l$, with $k_B$ being the Boltzman constant. The \cgls{fspl} is $L(k,i) = \left(4\pi f_c d(k, i) / c_0\right)^2$, where $f_c$ is the carrier frequency and $d(k, i)$ the distance between nodes $k$ and $i$ \cite{ippolito2017satellite, 9327501}. We assume fixed rate links operating at the minimum rate supported by the link. This is equivalent to selecting $d(k, i)$ as the maximum communication distance $d_\mathrm{Th}(k, i)$ between these nodes. Following \cite{leyva2022ngso}, we set  $f_c = \SI{20}{\giga\Hz}$, $B = \SI{500}{\mega\Hz}$, $P_t = \SI{40}{\dBm}$, $T = \SI{354}{\kelvin}$, and the antenna gains to \SI{32.13}{\dBi}.

Numerical \cgls{ml} performance is evaluated based on the two most widely used benchmarks. The first is a conventional $\num{7850}$-parameter logistic regression model trained on the MNIST dataset, $28 \times 28$ pixel greyscale images of handwritten digits ranging from 0 to 9 \cite{MNIST}.
The other is a deep CNN with \num{122570} parameters from \cite{fraboni2023general}, trained on the CIFAR-10 dataset \cite{Krizhevsky2009}, which includes 10 classes with $32 \times 32$ RGB images. The data samples are either evenly distributed at random among the satellites, which is the \cgls{iid} setting, or in a non-\cgls{iid} fashion using a Dirichlet distribution with parameter 0.5 \cite{yurochkin2019bayesian, wang2020federated}. We use a batch size of ten, run five local epochs, and take the learning rate as 0.1. A computation time $t_l$ of \SI{60}{\second} and \SI{480}{\second} is assumed for MNIST and CIFAR-10, respectively. The simulation is build upon the FedML framework \cite{he2020fedml}.

\subsection{Synchronous and Asynchronous Orchestration}
We start by evaluating the benefits of \cglspl{isl} for synchronous orchestration as detailed in \cref{alg:syncPS}. Without employing \cglspl{isl}, this is the vanilla \cgls{fl} approach as first proposed in \cite{mcmahan2017communication}.
Directly applying it to \cgls{sfl}, i.e., each satellite contacts the \cgls{ps} directly, leads to very slow convergence speed, especially in scenarios with terrestrial orchestration \cite{WCL_fedsat}. The question is whether the usage of \cglspl{isl} resolves the connectivity bottleneck sufficiently to make synchronous orchestration feasible.

To this end, we measure the test accuracy with respect to the wall clock time for synchronous terrestrial orchestration in the \Wstar and \Wdelta constellations, with and without \cglspl{isl}. We complement this by an experiment with a \cgls{ps} in \cgls{leo}. Results for non-\cgls{iid} MNIST and CIFAR-10 are displayed in {\color{black}\cref{fig:Sync_MNIST_CIFAR10}}.
All scenarios show the distinct step function behavior first observed in \cite{WCL_fedsat} for the non-\cgls{isl} scenarios. This is caused by the \cgls{ps} being forced to wait for all results before updating the global model. As expected, the usage of \cglspl{isl}, as proposed in \cref{sec:routing}, shortens the convergence time significantly in all scenarios.
With MNIST training usually showing very fast convergence, it can be easily observed that \cglspl{isl} reduce the time to convergence in all scenarios by 4 hours. This leads to almost instantaneous convergence in the \cgls{leo} scenario, at least in relation to the training duration in conventional \cgls{sfl}, and reduces the convergence time by up to \SI{50}{\percent} for terrestrial orchestration.

\begin{figure*}
	\centering
	\subfloat[\cgls{iid} distribution]{%
		\begin{tikzpicture}
	\begin{axis}[
yminorgrids = true, 
legend entries = {\cgls{isl} Async., \cgls{isl} Sync., non-\cgls{isl} Sync.},
xlabel={Time [h]},
ylabel={Test Accuracy},
				grid=major,
				minor x tick num = 4,
				minor y tick num = 1,
extra y ticks = {.772},
xmin = 0,
xmax = 110,
ymin = 0,
ymax = 0.79,
grid = major,
width=0.93*\axisdefaultwidth,
height=1*\axisdefaultheight,
legend cell align={left},
legend pos=south east,
legend style={font=\small}
]

\addplot[color=blue, line width = 1.1pt] table [x=Time_FedISL_Async_Bremen_IID, y=Acc_FedISL_Async_Bremen_IID, col sep=comma]{images/FedISL_Sync_Async_CIFAR10.csv};

\addplot[color=black!40!green,line width=1.1pt] table [x=Time_FedISL_Sync_Bremen_IID, y=Acc_FedISL_Sync_Bremen_IID, col sep=comma]{images/FedISL_Sync_Async_CIFAR10.csv};

\addplot[color=black!20!red,line width=1.1pt] table [x=Time_FedNonISL_Sync_Bremen_IID, y=Acc_FedNonISL_Sync_Bremen_IID, col sep=comma]{images/FedISL_Sync_Async_CIFAR10.csv};

			\end{axis}
		\end{tikzpicture}%
		\label{fig:Async_IID}%
	}
	\hfill%
	\subfloat[non-\cgls{iid} distribution]{%
		\begin{tikzpicture}
			\begin{axis}[
yminorgrids = true, 
legend entries = {\cgls{isl} Async., \cgls{isl} Sync., non-\cgls{isl} Sync.},
xlabel={Time [h]},
ylabel={Test Accuracy},
				grid=major,
				minor x tick num = 4,
				minor y tick num = 1,
extra y ticks = {.772},
xmin = 0,
xmax = 110,
ymin = 0,
ymax = 0.79,
grid = major,
width=0.93*\axisdefaultwidth,
height=1*\axisdefaultheight,
legend cell align={left},
legend pos=south east,
legend style={font=\small},
				yticklabel pos = right,
]

\addplot[color=blue, line width = 1.1pt] table [x=Time_FedISL_Async_Bremen_NIID, y=Acc_FedISL_Async_Bremen_NIID, col sep=comma]{images/FedISL_Sync_Async_CIFAR10.csv};

\addplot[color=black!40!green,line width=1.1pt] table [x=Time_FedISL_Sync_Bremen_NIID, y=Acc_FedISL_Sync_Bremen_NIID, col sep=comma]{images/FedISL_Sync_Async_CIFAR10.csv};

\addplot[color=black!20!red, line width = 1.1pt] table [x=Time_FedNonISL_Sync_Bremen_NIID, y=Acc_FedNonISL_Sync_Bremen_NIID, col sep=comma]{images/FedISL_Sync_Async_CIFAR10.csv};

			\end{axis}
		\end{tikzpicture}%
		\label{fig:Async_NIID}%
	}%
	\caption{\color{black} Comparison of synchronous and asynchronous orchestration. The figure displays test accuracy with respect to wall-clock time for a \Wdelta constellation with CIFAR-10 dataset, distributed \cgls{iid} and non-\cgls{iid}, and PS located in Bremen.}
 \label{fig:Async}
\end{figure*}
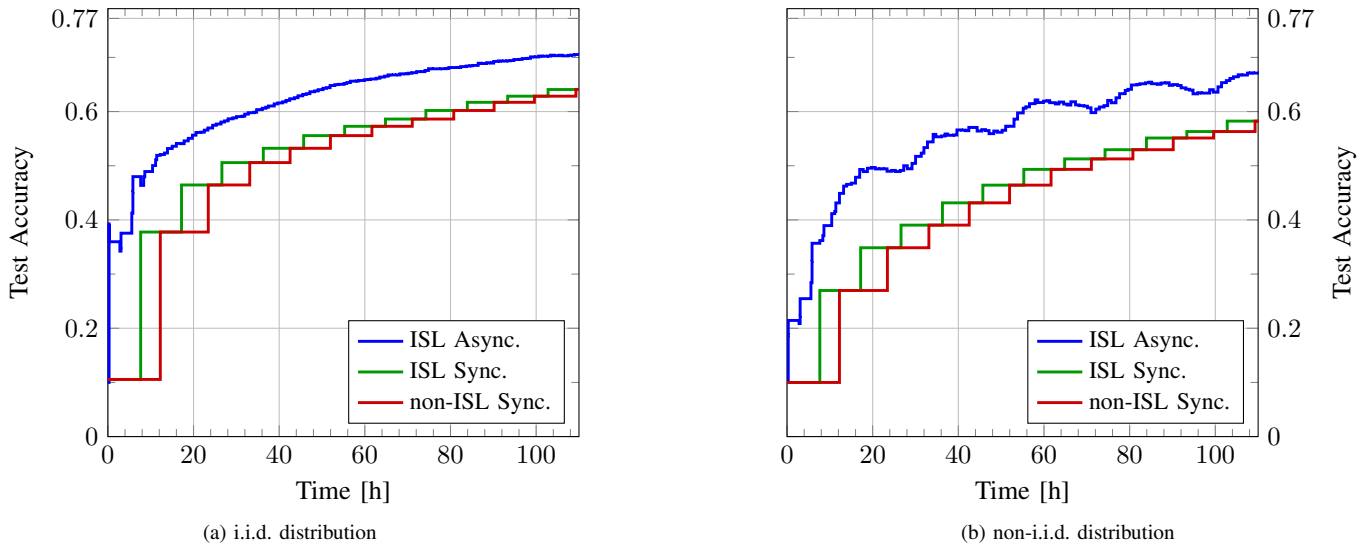

The CIFAR-10 experiment was chosen to evaluate the convergence behavior of a more involved \cgls{ml} model. As such, convergence requires a large number of global iterations. From \cref{fig:Sync_CIFAR10}, we observe a more nuanced convergence behavior as compared to the MNIST experiment. In particular, \cgls{leo}-based orchestration benefits most from \cglspl{isl}, while \Wdelta profits only in the initial learning phase. Surprisingly, it basically shows the same convergence speed as without \cglspl{isl}, simply shifted in time by a few hours. The reason behind this is best understood from \cref{fig:pattern}, which displays exactly the connectivity patterns of these two scenarios. Connectivity in the \cgls{leo} \cgls{ps} scenario changes from a sporadic towards a near-persistent connectivity pattern. Instead, for terrestrial orchestration, the connectivity windows grow significantly in duration, but the overall connectivity pattern is still sporadic. That is, the offline periods of individual orbital groups still lead to blocking in synchronous aggregation algorithms. However, this behavior is not universal to terrestrial orchestration. It depends on the constellation design and \cgls{gs} location, as can be seen by the improved convergence speed in the \Wstar scenario. Moreover, combining synchronous orchestration and \cglspl{isl} with the scheduling approach from \cite{razmi2022scheduling} is expected to alleviate this problem in training scenarios with limited computational complexity.

Following the discussion in \cref{sec:ps} and \cite{matthiesen2022federated}, a viable alternative to synchronous orchestration in sporadic connectivity scenarios is asynchronous \cgls{ps} operation. This is evaluated in \cref{fig:Async} for \Wdelta with terrestrial orchestration, with maximum time between updates $T_u$ set to 147 minutes. Convergence with asynchronous aggregation is improved over synchronous operation, but still involves at least a ten-fold increase in convergence time. Especially initial convergence speed, i.e., within the first few hours of training, is greatly increased. This leads to the conclusion that asynchronous aggregation is a potential solution if the system design suffers from sporadic connectivity. However, if changing the \cgls{ps} location is an option, it might be preferable to asynchronous orchestration.

\subsection{\color{black}{Failure Handling}}
{
To evaluate the performance of failure handling schemes, we consider single orbit with an inclination of 85° and altitude of 2000 km, where the \cgls{ps} is located in a \cgls{gs} in Bremen. We model the computation time for learning at any satellite $k$ as $T_l(k) = t_l(k) + X(k)$, where $t_l(k)$ is the deterministic time for learning, and $X(k)$ accounts for random additional computation time due to e.g. multiprocessing loads.}{\color{black} The distribution of $X(k)$ is modelled using a Gamma distribution, a common approach for modeling service times, with shape and scale parameters denoted as $\alpha(k)$ and $\theta(k)$ respectively \cite{Lorch2001}. Hence, the \cgls{cdf} of $X(k)$ is
$F_{X}(x) = \Gamma(\alpha(k))^{-1}\; \gamma(\alpha(k), \frac{x}{\theta(k)})$ if $x$
is positive, and $F_{X}(x) = 0$ otherwise, where $\alpha(k)> 0$, $\theta(k) > 0$,  $\Gamma(z)$ is the Gamma function, and $\gamma(s, x)$ is the lower incomplete Gamma function. We set the values of $t_l(k)$, $\alpha(k)$, and $\theta(k)$ for any satellite $k$ to 480 \si{seconds}, 25, and 25 \si{seconds} respectively.
The random communication delays in any satellite $k$, denoted as $Y(k)$, are taken into account in addition to the deterministic time $t_c(k,j)$ for transmitting the aggregated parameters from satellites $k$ to $j$. In this regard, total communication time for each satellite $k$ is modelled as $T_c(k,j) = t_c(k,j) + Y(k)$, where $Y(k)$ follows an exponential distribution with parameter $\lambda(k) > 0$, with \cgls{cdf}
$F_{Y}(y) = 1- \exp(-y \lambda(k))$ if $y$ is positive and $F_{Y}(y)=0$ otherwise. Here, we set $t_c(k,j)$ and $\lambda(k)$ for any satellite $k$ to 50 \si{seconds} and \num{0.025} respectively. It is worth mentioning that, when the delays in the transmissions and learning procedures of satellites are very small, the system may not need any failure handling scheme. In this regard, we have set the parameters values such that the effect of failure can be seen. We also set $t_{g}$ to 0.}\\ 
{\color{black} \Cref{{fig:FH schemes}} shows the required time for handling the failure with respect to the number of satellites in the orbit, for the pass-to-neighbor and determine-new-sink schemes.
For any number of satellite in \cref{{fig:FH schemes}}, we calculate the average required time based on overall \num{20000} random values for $X(k)$ and $Y(k)$ for the 20 different randomly chosen global iterations, 
 starting at different times. As we can see in \cref{{fig:FH schemes}}, with the increase in the number of satellites, the determine-new-sink scheme reduces the required time by a factor of approximately $4.5$, as compared to the pass-to-neighbor in any orbit. This is particularly effective when considering the delays incurred in constellations with multiple orbits, as the determine-new-sink failure handling scheme significantly reduces the required time within each orbit, thereby substantially reducing the overall required time.}

 {
\begin{figure}
\centering
    		\begin{tikzpicture}
	\begin{axis}[
 font = {\small},
yminorgrids = true, 
legend entries = {determine-new-sink, pass-to-neighbor},
xlabel={Number of satellites in an orbital plane},
ylabel={Time for failure handling [s]},
				grid=major,
				minor x tick num = 4,
				minor y tick num = 1,
extra y ticks = {0},
xmin = 0,
xmax = 52,
ymin = 0,
ymax = 1600,
grid = major,
width=0.93*\axisdefaultwidth,
height=1*\axisdefaultheight,
legend cell align={left},
legend pos= north west,
legend style={font=\small}
]
\addplot[color=blue, line width = 1.1pt] table [x=Num_satellites, y=Proposed_FH, col sep=comma]{images/FH_schemes.csv};

\addplot[color=black, line width = 1.1pt] table [x=Num_satellites, y=Clock_wise_FH, col sep=comma]{images/FH_schemes.csv};

			\end{axis}
		\end{tikzpicture}%
    \caption{The time required for failure handling (FH) with respect to the number of satellites in an orbital plane. }
    \label{fig:FH schemes}
\end{figure}
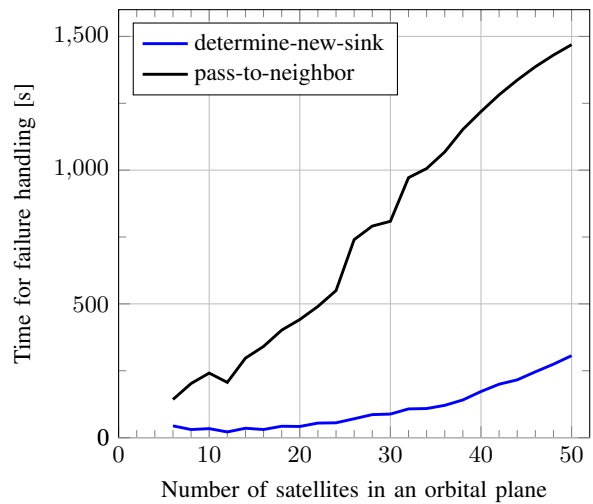}

\subsection{Gradient Sparsification and Transmission Load}

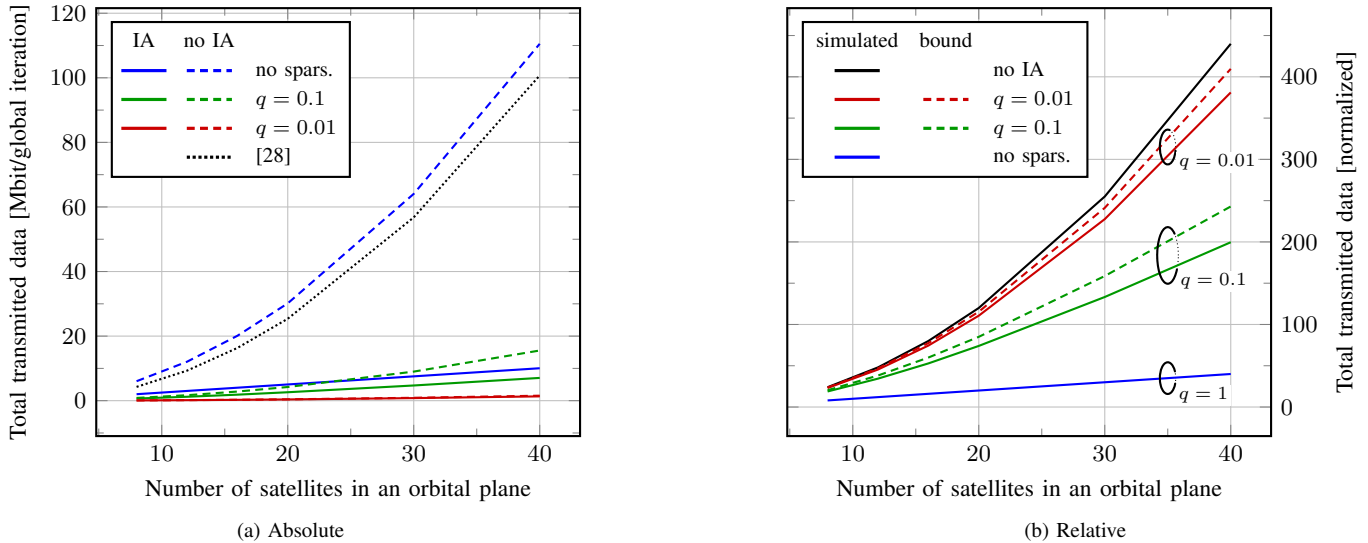
\begin{figure*}
	\centering%
	\subfloat[Absolute]{%
		\label{fig: Total transmitted data Absolute}%
		\begin{tikzpicture}
			\begin{axis} [
					thick,
					font = {\small},
					xlabel={\color{black} Number of satellites in an orbital plane},
					ylabel={Total transmitted data [\si{\mega\bit}/global iteration]},
					ylabel near ticks,
					grid=major,
					minor x tick num = 1,
					minor y tick num = 1,
					scaled y ticks = base 10:-6,
					ytick scale label code/.code={},
					no markers,
					legend columns = 3,
					legend cell align=left,
					legend pos = {north west},
					legend style = {font={\footnotesize\mystrut}},
					legend image post style = {very thick},
					width=.95*\axisdefaultwidth,
					height=\axisdefaultheight,
				]

				\pgfplotstableread[col sep=comma]{images/totaldata_IA_updated_version_other_references.dat}\tbl

				\addlegendimage{legend image with text=IA}
				\addlegendentry{}
				\addlegendimage{legend image with text=no IA}
				\addlegendentry{}
				\addlegendimage{empty legend}
				\addlegendentry{}

				\addplot[blue] table [x=sats, y=simulated_1] {\tbl};
				\addlegendentry{}
				\addplot[densely dashed, blue] table [x=sats, y=no_IA_1] {\tbl};
				\addlegendentry{}
				\addlegendimage{empty legend}
				\addlegendentry{no spars.}

				\addplot[black!40!green] table [x=sats, y=simulated_0.1] {\tbl};
				\addlegendentry{}
				\addplot[densely dashed, black!40!green] table [x=sats, y=no_IA_0.1] {\tbl};
				\addlegendentry{}
				\addlegendimage{empty legend}
				\addlegendentry{$q = 0.1$}

				\addplot[black!20!red] table [x=sats, y=simulated_0.01] {\tbl};
				\addlegendentry{}
				\addplot[densely dashed, black!20!red] table [x=sats, y=no_IA_0.01] {\tbl};
				\addlegendentry{}
				\addlegendimage{empty legend}
				\addlegendentry{$q = 0.01$}

                   \addlegendimage{empty legend}
                   \addlegendentry{}
				\addplot[black, densely dotted] table [x=sats, y=Elmahallawy] {\tbl};
                    \addlegendimage{empty legend}
                    \addlegendentry{}
				\addlegendentry{\cite{elmahallawy2023optimizing}}
                    \addlegendimage{empty legend}

			\end{axis}
	\end{tikzpicture}}%
	\hfill%
	\subfloat[Relative]{%
		\label{fig: Total transmitted data Relative}%
		\begin{tikzpicture}
			\begin{axis} [
					thick,
					font = {\small},
					ylabel style = {align=center},
					xlabel={\color{black}Number of satellites in an orbital plane},
					ylabel={Total transmitted data [normalized]},
					grid=major,
					minor x tick num = 1,
					minor y tick num = 1,
					no markers,
					legend columns = 3,
					legend cell align=left,
					legend pos = {north west},
					legend style = {font={\footnotesize\mystrut}},
					legend image post style = {very thick},
					yticklabel pos = right,
					width=.95*\axisdefaultwidth,
					height=\axisdefaultheight,
				]

				\pgfplotstableread[col sep=comma]{images/IA_vs_q.dat}\tbl

				\coordinate (b001) at (axis cs:35,325.5);
				\coordinate (s001) at (axis cs:35,304.4);
				\coordinate (b01) at (axis cs:35,200.8);
				\coordinate (s01) at (axis cs:35,166.5);
				\coordinate (b1) at (axis cs:35,35);

				\draw[line width = .4pt, densely dotted] let \p1 = (b001), \p2 = (s001), \n1 = {veclen((\x1-\x2, \y1-\y2))} in ($(b001)!.5!(s001)$) ellipse [y radius = \n1, x radius = 3pt];
				\draw[line width = .4pt, densely dotted] let \p1 = (b01), \p2 = (s01), \n2 = {veclen((\x1-\x2, \y1-\y2))}, \n1 = {4pt} in ($(b01)!.5!(s01)$) ellipse [y radius = \n2, x radius = \n1];
				\draw [line width = .4pt, densely dotted](b1) ellipse [y radius = 6pt, x radius = 3pt];

				\addlegendimage{legend image with text=simulated}
				\addlegendentry{}
				\addlegendimage{legend image with text=bound}
				\addlegendentry{}
				\addlegendimage{empty legend}
				\addlegendentry{}

				\addplot[] table [x=sats, y=noIA] {\tbl};
				\addlegendentry{}
				\addlegendimage{empty legend}
				\addlegendentry{}
				\addlegendimage{empty legend}
				\addlegendentry{no IA}

				\addplot[black!20!red] table [x=sats, y=simulated_0.01] {\tbl};
				\addlegendentry{}
				\addplot[densely dashed, black!20!red] table [x=sats, y=bound_0.01] {\tbl};
				\addlegendentry{}
				\addlegendimage{empty legend}
				\addlegendentry{$q = 0.01$}

				\addplot[black!40!green] table [x=sats, y=simulated_0.1] {\tbl};
				\addlegendentry{}
				\addplot[densely dashed, black!40!green] table [x=sats, y=bound_0.1] {\tbl};
				\addlegendentry{}
				\addlegendimage{empty legend}
				\addlegendentry{$q = 0.1$}

				\addplot[blue] table [x=sats, y=IA] {\tbl};
				\addlegendentry{}
				\addlegendimage{empty legend}
				\addlegendentry{}
				\addlegendimage{empty legend}
				\addlegendentry{no spars.}

				\draw[line cap = round] let \p1 = (b001), \p2 = (s001), \n2 = {veclen((\x1-\x2, \y1-\y2))}, \n1 = {3pt} in ($(b001)!.5!(s001)$) + (76:\n1 and \n2) arc (76:348:\n1 and \n2) node[anchor=north west,font=\scriptsize, inner sep = 1pt, fill = white] {$q = 0.01$};
				\draw[line cap = round] let \p1 = (b01), \p2 = (s01), \n2 = {veclen((\x1-\x2, \y1-\y2))}, \n1 = {4pt} in ($(b01)!.5!(s01)$) + (50:\n1 and \n2) arc (50:330:\n1 and \n2) node[anchor=north west,font=\scriptsize, inner sep = 1pt, fill = white] {$q = 0.1$};
				\draw[line cap = round] (b1) + (25:3pt and 6pt) arc (25:340:3pt and 6pt) node[anchor=north west,font=\scriptsize, inner sep = 1pt, fill = white] {$q = 1$};
			\end{axis}
	\end{tikzpicture}}%
     \caption{Total transmitted data {\color{black} with respect to the number of satellites in an orbit per} global iteration for three different sparsification ratios; a) $q=0.01$ and b) $q=0.1$, and c) without sparsification (q=1). IA and no IA stand for the proposed algorithm with and without incremental aggregation respectively.}
     \label{fig:Communication_Load}
\end{figure*}

\Cref{fig:Communication_Load} shows the transmission load required for collecting the gradients in one orbit and sending them towards the PS, within a single global iteration. The results are based {\color{black} on running FedAvg} on the MNIST dataset with non-\cgls{iid} distribution. The figure depicts the total transmitted data {\color{black} with and without incremental aggregation (IA)} as a function of the {\color{black} number of satellites within an orbital plane. We consider three different cases, without sparsification as well as sparsification with ratios $q = 0.1$ and $q = 0.01$.}
When IA is used, each satellite aggregates its gradient updates with the received ones, according to the method described in \cref{sec:routing:agg}. Without IA, each satellite only forwards its gradients and the received ones to the other satellite or the \cgls{ps}, without performing any aggregation.

\begingroup
\color{black}
\Cref{fig: Total transmitted data Absolute} shows the communication effort in terms of the total transmitted data. We set the storage size $\omega$ of model parameters to 32\,bit. The sparsified vectors need an additional index field per \color{black}entry, which is chosen to be $13$\,bits. Hence, each nonzero element in a sparsified vector requires a total of $45$\,bits. The effect of sparsification on transmission cost is well known and not further discussed here. A notable observation is that IA has comparable (and even stronger) effect than removing 90\,\% of the data from each vector (with sparsification). This effect continues to hold in the combination of sparsification and IA, leading to an even further reduction of communication cost and, more importantly, a linear cost increase in the number of satellites.

In addition, we compare the communication efficiency of our approach to the state-of-the-art in \cite{elmahallawy2023optimizing}. There, each satellite within an orbital plane transmits its updated local parameters to a designated sink satellite through multihop connections over neighbouring satellites. In contrast to our approach, each parameter vector is transmitted through unicast transmissions as in the baseline approach. However, contrary to the baseline, the sink satellite performs partial aggregation before transmitting the updated parameter vector to the \cgls{ps}. The result is a quadratic growth in total communication effort, with slightly better performance than the baseline. As the number of satellites per orbit increases, we observe a tenfold reduction in communication effort due to IA over \cite{elmahallawy2023optimizing} in \cref{fig: Total transmitted data Absolute}.

The impact of IA appears to decrease in \cref{fig: Total transmitted data Absolute} with stricter sparsification, i.e., the effect of IA seems small for $q = 0.1$ and negligible for $q = 0.01$. This, however, is only partially true and near impossible to assess from \cref{fig: Total transmitted data Absolute}. For this reason, \cref{fig: Total transmitted data Relative} shows the total transmitted data normalized to the size of a single parameter vector ready for transmission. The purpose of this is to evaluate the increase in communication efficiency due to IA independently from the compression achieved by sparsification. The normalized baseline, i.e., multiple unicast transmission without IA, is identical with and without sparsification and, hence, only shown once. Based on this normalization, the impact of IA on the communication efficiency is proportional to the size of the gap between the simulated result (solid color) and the black baseline. The initial impression from \cref{fig: Total transmitted data Absolute} continues to hold, i.e., the efficiency of IA decreases with increasing sparsification ratio, i.e., lower $q$. However, this does not imply IA is expendable in any way. For 40 satellites per orbital plane, we observe a reduction in communication cost of 55\,\% for $q = 0.1$ purely due to IA. Without sparsification, this reduction amounts to 91\,\%, while it is still a notable 13\,\% for $q = 0.01$. An important aspect to consider is that a sparsification ratio of $q = 0.01$ leads to slower convergence in training, as opposed to $q = 0.1$, which might deteriorate the savings in communication cost. Instead, IA comes at no cost to the training process and, thus, can mitigate some of the adverse effects that the lossy compression (sparsification) has on a training process within a fixed communications budget.

A possible cause for the reduced efficiency of IA at small $q$ is indicated by the dashed lines in \cref{fig: Total transmitted data Relative}. These display the estimated communication effort based on the bound in \cref{prop:sparse} and \cref{eq:sparseeff}. We observe that this bound becomes less tight as the number of satellites increases and as the vectors become sparser.\footnote{Recalling the discussion in \cref{sec:sparsification}, this has no direct impact on the communication effort and should only lead to small timing errors in the predictive routing procedure.} This discrepancy between bound and simulation is most likely caused by the fact that the computed local updates are correlated instead of being stochastically independent as assumed in \cref{prop:sparse}. In other words, a large gap between bound and simulation indicates a stronger correlation between the position of the nonzero elements in the gradients, while a small gap points to their distribution being closer to independence. Now, with higher sparsification ratios (small $q$), the probability of two vectors having overlapping nonzero entries becomes increasingly small and, hence, each IA step increases the number of nonzero entries in the resulting vector significantly. To see how this reduces the effect of IA, recall that a sparse vector is represented by the nonzero values together with the indices of these nonzero positions. 
When the indices of non-zero elements differ among satellites, the transmitted vector after IA must contain all nonzero elements of the received vector together with the nonzero elements of the satellite's own sparsified result. With decreasing overlap between the positions of nonzero elements, this becomes closer to separate transmission of both vectors.

\Cref{fig:Communication_Load} deals with the amount of total transmitted data necessary to deliver all gradient updates from a client cluster towards the \cgls{ps}. We have observed that conventional approaches scale quadratically in the number of satellites per cluster. Instead, the proposed IA method scales linearly and, thus, offers a massive increase in communication efficiency. However, through the narrow lens of communication efficiency in terms of total transmitted data, the same efficiency is achieved by not using clusters at all and, instead, have each satellite communicate directly with the \cgls{ps}. To take a slightly different angle, recall that the majority of transmissions for clustered \cgls{sfl} is done over stable \cglspl{isl}, while the direct satellite-to-\cgls{ps} approach requires more costly \cgls{gsl} or a dynamic long-range \cgls{isl}. Consequently, the proposed framework reduces the communication load at the \cgls{ps}, in comparison to direct connectivity, directly proportional to the cluster size.

\endgroup

\section{Discussion and Conclusions} \label{sec: Discussion and Conclusions}

We have designed a clustered \cgls{fl} system tailored to distributed \cgls{ml} in modern satellite megaconstellations. It efficiently uses {\color{black}intra-orbit} \cglspl{isl} to avoid connectivity bottlenecks that impair convergence speed. Owing to a \cgls{fl}-specific in-network aggregation strategy, this is achieved without increasing the total transmitted amount of data. Indeed, it can be expected that this method actually decreases the communication cost due to relying on cheaper-to-operate \cglspl{isl} instead of long distance \cglspl{gsl}. Moreover, the communication load is evenly spread among the network instead of focusing all communications on direct worker-to-\cgls{ps} links.
This strategy requires a 
careful predictive route planning
in order to operate successfully in a dynamic network topology imposed by orbital mechanics. We have developed the necessary routing algorithms for this in conjunction with a rigorous distributed system design that inflicts a low overhead for synchronization and consensus finding.
The proposed system is compatible with a wide variety of \cgls{fl} algorithms, supports gradient sparsification, and includes facilities for asynchronous clustered aggregation. While not explicitly mentioned, the system's extension to incorporate worker scheduling is rather straightforward. 
We have evaluated the performance of the proposed system for a few carefully chosen examples. The results highlight the major increase in convergence speed {\color{black}to} \cglspl{isl} and the bandwidth effectiveness of our in-network aggregation approach. 

Regarding future work, it is interesting to revise the assumption adopted in this paper that
all participants in the \cgls{fl} process are trustworthy, under complete control, and the software operates nominally. For instance, if the satellites belong to different operators, the use of ISLs may incur a certain cost and/or a group for satellites may decide not to participate in the FL process, which would require an extension of the proposed algorithms.
{\color{black} Moreover, incorporating inter-orbit \cglspl{isl} in the communication scheme has the potential to replace out-of-constellation orchestration and further increase the resource-efficiency of \cgls{sfl}. This, however, comes at the cost of higher management complexity, will likely require investigating several special cases, e.g., with and without cross-seam \cglspl{sfl}, and opens the possibility for \cgls{sfl}-specific network topology design.} {\color{black}It is also noteworthy that our scenario with one \cgls{gs} can be extended to the case with multiple \cglspl{gs} to improve the satellite-\cgls{ps} connectivity. Two principal approaches to employ multiple \cglspl{gs} as \cgls{ps} can be considered: (1) the \cglspl{gs} forward the communication towards a centralized cloud \cgls{ps}, or (2) they act together as a distributed \cgls{ps}. The first case is incremental with respect to the proposed algorithms as it involves multi-hop routing through a terrestrial network. The second case implies a stronger separation of ground- and space-segment and requires accurate synchronization to maintain model parameters integrity.}

\balance
\bibliography{IEEEtrancfg,references.bib}

\end{document}